\documentclass[%
 reprint,
superscriptaddress,
 amsmath,amssymb,
 aps,
]{revtex4-1}

\usepackage{pgfplots}
\pgfplotsset{/pgf/number format/use comma,compat=newest}
\usepackage{graphicx}
\usepackage{graphics}
\usepackage{epstopdf}
\usepackage{dcolumn}
\usepackage{bm}

\usepackage{tikz} 
\usepackage{bbold}
\usepackage[extra]{tipa}
\usepackage{float}
\usepackage{color}
\usepackage[colorlinks=true,citecolor=red,linkcolor=blue]{hyperref}
\usepackage{soul}

\DeclareMathOperator{\arcsinh}{arcsinh}

\newcommand{\ket}[1]{|{#1}\rangle}                       
\newcommand{\bra}[1]{\langle {#1}|}                      
\newcommand{\average}[1]{\langle {#1} \rangle}           

\newcommand{\ketbra}[2]{\left\vert#1\right\rangle\left\langle#2\right\vert}

\begin{document}

\preprint{APS/123-QED}

\title{Exactly solvable time-dependent models of two interacting two-level systems}

\author{R. Grimaudo}
\address{%
 Dipartimento di Fisica e Chimica dell'Universit\`a di Palermo, Via Archirafi, 36, I-90123 Palermo, Italy.
}
\author{A. Messina}%
\address{%
 Dipartimento di Fisica e Chimica dell'Universit\`a di Palermo, Via Archirafi, 36, I-90123 Palermo, Italy.
}%
\address{I.N.F.N., Sezione di Catania}

\author{H. Nakazato}
\affiliation{
 Department of Physics, Waseda University, Tokyo 169-8555, Japan\\
}%


\date{\today}

\begin{abstract}
Two coupled two-level systems placed under external time-dependent magnetic fields are modelled by a general Hamiltonian endowed with a symmetry that enables us to reduce the total dynamics into two independent two-dimensional sub-dynamics.  Each of the sub-dynamics is shown to be brought into an exactly solvable form by appropriately engineering the magnetic fields and thus we obtain an exact time evolution of the compound system.  Several physically relevant and interesting quantities are evaluated exactly to disclose intriguing phenomena in such a system.

\begin{description}
\item[PACS numbers]
03.65.Aa, 0.3.67.Ac, 07.05.Dz, 75.10.Jm
\item[Keywords]
Solvable model, Quantum two level system, Time-ependent Hamiltonian, Quantized spin models.
\end{description}
\end{abstract}

\maketitle


\section{Introduction}
A rigid and localized dimeric structure (simply dimer) consists of a pair of independent distinguishable quantum subsystems living, by definition, in finite-dimensional Hilbert spaces $\mathcal{H}_1$ and $\mathcal{H}_2$ and, therefore, hereafter referred to as spins $\hat{\mathbf{S}}_1 \equiv (\hat{S}_1^x,\hat{S}_1^y,\hat{S}_1^z)$ and $\hat{\mathbf{S}}_2 \equiv (\hat{S}_1^x,\hat{S}_1^y,\hat{S}_1^z)$ respectively, $\hat{S}_i^a$ $(i=1,2;\;a= x,y,z)$ being the operator for the $a$-cartesian component of $\hat{\mathbf{S}}_i$ in the laboratory reference frame. The dimension of the Hilbert space $\mathcal{H} = \mathcal{H}_1 \otimes \mathcal{H}_2$ of the dimer is $(2S_1 +1)(2S_2+1)$, indeed postulating the absence in the two subsystems as well as in the compound system of classical degrees of freedom (situation previously described using the adjectives `rigid' and `localized').
The physical nature of $\hat{{\bm S}}_i$ depends on the particular scenario under scrutiny: it may be the spin of an electron or a nucleus, the angular momentum of an atom in its ground state or an effective representation of a few-level system dynamical variable.
The Hamiltonian $H$ of the dimer is then a true or effective spin Hamiltonian where the terms linear in $\hat{S}_i^a$ may (even fictitiously) be interpreted as Zeeman coupling of each of the two spins with classical, external, generally different and time-dependent effective magnetic fields $\mathbf{B}_1(t)$ and $\mathbf{B}_2(t)$ while the bilinear contributions may be thought of as stemming from the spin-spin interaction \cite{Bolton}.

Over the last two decades a great deal of theoretical, experimental and applicative attention has been devoted to the field of Molecular Magnetic Materials, in particular after the discovery of the so-called Single Magnet Molecule (SSM), that is a single molecule behaving like a nanosized magnet associated to an unusual high value (even $S=10$ \cite{Thomas}) of the spin in the ground state of the molecule.
It is a matter of fact that as a result of a successful, extraordinary and synergically interdisciplinary effort aimed at searching and producing SMM in laboratory, in the last few years we have witnessed a very fast growing of efficient protocols for synthesizing a variety of such molecular magnets with the added value of possessing a number of constituent paramagnetic ions embodied in the molecule running from 2 to 10 in different samples \cite{Calbucci}.
Such  important technological advances on the one hand open very good applicative perspectives in many directions, from the realization of an experimental set up for testing theoretical prediction concerning qudits-based single purpose quantum computers to the availability of new materials with magnetic properties tailored on demand to meet specific tasks. On the other hand the production of crystalline or powder samples made up of molecular magnetic units, provides an ideal platform to investigate and reveal the emergence of nonclassical signatures in the quantum dynamics of two or few interacting spins.

The simplest coupled spin system we may conceive consists, of course, of two interacting spin 1/2's only in a dimer, isolated from its environment (rest of the sample) degrees of freedom.
Some binuclear copper(II) compounds, e.g.~\cite{Napolitano} \cite{Calvo}, provide a possible scenario of this kind and in the previous references the values of the parameters characterizing the spin-spin interaction in such a molecule have been experimentally determined exploiting electron-paramagnetic resonance techniques. Motivations to investigate the emergence of quantum signatures in the behaviour of two coupled spins ($\geq 1/2$) go beyond the area of magnetic materials. Two spin 1/2 Hamiltonians provide indeed experimentally implementable powerful effective models to capture quantum properties of such systems like two coupled semiconductor quantum dots \cite{Bagrov} or a pair of two neutral cold atoms each nested into two adjacent sites of an optical lattice made up of an isolated double wells \cite{Anderlini}.
Spin models provide a successful language to investigate possible manipulations of the qubits aimed at quantum computing purposes and quantum information transfer between two spin-qubits \cite{Nguyen}, encompassing rather different physical contents like, for example, cavity QED \cite{Imamoglu} \cite{Zheng}, superconductors \cite{Wang} \cite{Arnesen} and trapped ions \cite{Yi} \citep{PorrasCirac}. 

The most general Hamiltonian model of an isolated dimer hosting two spin 1/2's may be written as a bilinear form involving the two sets of operators $\{ \hat{S}_1^x, \hat{S}_1^y, \hat{S}_1^z, \hat{S}_1^0 \}$ and $\{ \hat{S}_2^x, \hat{S}_2^y, \hat{S}_2^z, \hat{S}_2^0 \}$, that is,
\begin{equation} \label{General Hamiltonian}
H = \sum_{(i,j) \neq (0,0)} \gamma_{ij} \hat{S}_1^i \otimes \hat{S}_2^j
\end{equation}
where $i$ and $j$ run in the set ($x,y,z,0$) and the operator $\hat{S}_i^0$ ($i=1,2$) is the identity operator $\openone_i$ in $\mathcal{H}_i$. The six real parameters $\gamma_{i0}$ and $\gamma_{0j}$ ($i \cdot j \neq 0$) are assumed to be generally time dependent while all the other parameters characterizing the spin-spin coupling are real and time independent. Without further specific constraints on the 15 parameters $\gamma_{ij}$ $(i,j = x,y,z,0)$, the Hamiltonian possesses no symmetries and in particular it does not commute with the collective angular momentum operators $\hat{\mathbf{S}}^2 = (\hat{\mathbf{S}}_1 + \hat{\mathbf{S}}_2)^2$ and/or $\hat{S}^z = \hat{S}_1^z + \hat{S}_2^z$. In such a case, even if $H$ is time independent, the four roots of the relative secular equation, albeit determinable, are rather involved functions of all the 15 parameters and then are practically not exploitable for extracting physical prediction on the physical system under scrutiny. Thus, either legitimated by investigations on specific physical situations or motivated by the interest in studying models possessing, by construction, constants of motion, some constraints on the parameters $\gamma_{ij}$ have been introduced in the literature, making the Hamiltonian \eqref{General Hamiltonian} less general and at the same time non trivial and of physical interest. It is enough to quote the main declinations of the three-dimensional quantum Heisenberg models or the Dzyaloshinskii-Moriya (DM) models \cite{Albayrak} \cite{Guerrero} in conjugation or not with simplified contributions to terms describing anisotropy effects in the Hamiltonian.

In this paper we too investigate a Hamiltonian model included in \eqref{General Hamiltonian}, still general enough to remain not commutating with $\hat{\mathbf{S}}^2$ and $\hat{S}^z$, but such to posses a symmetry property at the origin of significant properties characterizing its quantum dynamics. A peculiar aspect of such a symmetry property is that it displays its usefulness even when we wish to study our physical system in a time-dependent scenario. Exploiting, indeed, the symmetry-induced reduction of the quantum dynamics generated by the time-dependent Hamiltonian we are going to propose to two dynamically invariant proper subspaces of $\mathcal{H}$, we are able to successfully apply a recently reported \cite{Mess-Nak} systematic approach for generating exactly solvable quantum dynamics of a single spin 1/2 subjected to a time-dependent magnetic field. Thus the main result of this paper is twofold. First we report the exact explicit solution of the time-dependent Schr\"odinger equation of a system of two coupled spin 1/2's described by a time-dependent generalized Heisenberg model. Second, we demonstrate that the method reported in Ref.~\cite{Mess-Nak}, even as it stands, proves to be a useful tool to treat more complex time-dependent scenarios.

The paper is organized as follows. The Hamiltonian model and the decoupling procedure are discussed in Section \ref{Section II}, where, in addition, the structure of the time-evolution operator is also constructed with the help reported in Ref.~\cite{Mess-Nak}. In the subsequent Section \ref{Section III}, some exactly solvable time-dependent Hamiltonian models of the two coupled qubits are singled out and analysed. Sections \ref{Section IV} and \ref{Section V} are respectively dedicated to a systematic study of the time behaviour of exemplary collective spin operators and of the concurrence. Some conclusive remarks are finally reported in the last section \ref{Section VI}.

\section{The Hamiltonian Model}\label{Section II}
By construction, the Hamiltonian model \eqref{General Hamiltonian} includes all possible contributions stemming from internal or external couplings of our two spin 1/2 system. It may be suggestively cast in the following form
\begin{equation} \label{Mimic Hamiltonian}
\begin{aligned}
{H'} = 
\mu_{B} ( \mathbf{B}_{1} \cdot \mathbf{g}_{1} \cdot \mathbf{S}_{1} + \mathbf{B}_{2} \cdot \mathbf{g}_{2} \cdot \mathbf{S}_{2})
+ \mathbf{S}_{1} \cdot \mathbf{\Gamma}_{12} \cdot \mathbf{S}_{2},
\end{aligned}
\end{equation}
where $\mathbf{g}_{1}$, $\mathbf{g}_{2}$ and $\mathbf{\Gamma}_{12}$ are appropriate second-order cartesian tensors whose entries are related to the 15 parameters appearing in Eq.~\eqref{General Hamiltonian} and $\mu_{B}$ denotes the Bohr magneton.
Equation \eqref{Mimic Hamiltonian} mimics the usual way of representing the Hamiltonian used in a molecular or nuclear context to describe the coupling of two true spin 1/2's.
In general we may claim that $\mathbf{g}_{1}$ and $\mathbf{g}_{2}$ include possible corrections to the coupling terms between each spin and its local time-dependent external magnetic field, while the other term includes contact term-like couplings as well as anisotropic-like spin-spin couplings.

The model we are going to propose is based on the following assumptions: a) $\mathbf{B}_1(t)$ and $\mathbf{B}_2(t)$ are at any time directed along the $\hat{z}$-axis of the laboratory frame, namely $\mathbf{B}_{i}(t) \equiv (0,0,B_{i}^{z}(t))$ ($i=1,2$); b) the cartesian tensor $\mathbf{\Gamma}_{12}$ has the following form
\begin{equation}
\mathbf{\Gamma}_{12}=
\begin{pmatrix}
\gamma_{xx} & \gamma_{xy} & 0 \\
\gamma_{yx} & \gamma_{yy} & 0 \\
0 & 0 & \gamma_{zz}
\end{pmatrix};
\end{equation}
c) the $\mathbf{g}$-tensors have the following form
\begin{equation}
\mathbf{g}_{i}=
\begin{pmatrix}
g_{i}^{xx} & g_{i}^{xy} & 0 \\
g_{i}^{yx} & g_{i}^{yy} & 0 \\
0 & 0 & g_{i}^{zz}
\end{pmatrix}
\end{equation}
($i=1,2$). The structure of the previous tensors is, for example, appropriate when the dimer coincides with a binuclear unit characterized by a $C_2$-symmetry with respect to the $\hat{z}$ axis as in Ref.~\cite{Calvo}.

In accordance with our previous assumptions, in this paper we investigate the quantum dynamics of the following time-dependent two spin Hamiltonian model
\begin{equation} \label{Hamiltonian}
\begin{aligned}
{H} = 
&\hbar\omega_{1}\hat{\sigma}_{1}^{z}+\hbar\omega_{2}\hat{\sigma}_{2}^{z}+\\
&+\gamma_{xx}\hat{\sigma}_{1}^{x}\hat{\sigma}_{2}^{x}+\gamma_{yy}\hat{\sigma}_{1}^{y}\hat{\sigma}_{2}^{y}+\gamma_{zz}\hat{\sigma}_{1}^{z}\hat{\sigma}_{2}^{z}+\\
&+\gamma_{xy}\hat{\sigma}_{1}^{x}\hat{\sigma}_{2}^{y}+\gamma_{yx}\hat{\sigma}_{1}^{y}\hat{\sigma}_{2}^{x},
\end{aligned}
\end{equation}
where $\hat{\sigma}_{i}^{x}$, $\hat{\sigma}_{i}^{y}$ and $\hat{\sigma}_{i}^{z}$ ($i=1,2$) are the Pauli matrices related to the respective components of the spin operator $\hat{\mathbf{S}}_{i}$ as
\begin{equation}
\hat{\mathbf{S}}_{i} = \dfrac{\hbar}{2} \hat{\bm{\sigma}}_{i}
\end{equation}
with $\hat{\bm{\sigma}}_{i} \equiv (\hat{\sigma}_{i}^{x},\hat{\sigma}_{i}^{y},\hat{\sigma}_{i}^{z})$, while
\begin{equation}\label{Relation omega-B}
\omega_{i}(t) = \dfrac{\mu_{B} g_{i}^{zz} B_{i}^{z}(t)}{2}.
\end{equation}
Note that the identity operators $\openone_i$ are and will mostly be suppressed for notational simplicity.

\subsection{Symmetry-based decoupling of the two spins}
As anticipated, our Hamiltonian does not commute with $\hat{\mathbf{S}}^2$ and $\hat{S}^z$ but it has been constructed in such a way to exhibit the following canonical and symmetry transformation
\begin{equation}\label{Symmetry Canonical Transformation}
\hat{{\sigma}}_{i}^{x}\to-\hat{\sigma}_{i}^{x},\quad
\hat{{\sigma}}_{i}^{y}\to-\hat{\sigma}_{i}^{y},\quad
\hat{{\sigma}}_{i}^{z}\to\hat{\sigma}_{i}^{z},\quad i=1,2.
\end{equation}
This fact implies the existence of a unitary time-independent operator accomplishing the transformation \eqref{Symmetry Canonical Transformation}, which is by construction a constant of motion.
It is easy to convince oneself that this unitary operator is given by $\pm \hat{\sigma}_{1}^{z} \hat{\sigma}_{2}^{z}$, being the transformation \eqref{Symmetry Canonical Transformation} nothing but the rotations of $\pi$ around the $\hat{z}$ axis with respect to each spin. Indeed, we can write the unitary operator accomplishing this transformation as follows
\begin{equation}\label{Rotation Operator}
e^{i \pi \hat{S}_{1}^{z} / \hbar} \otimes e^{i \pi \hat{S}_{2}^{z} / \hbar}
= - \hat{\sigma}_{1}^{z} \hat{\sigma}_{2}^{z} = \cos \Bigl( {\pi \over 2} \hat{\Sigma}_{z} \Bigr),
\end{equation}
where we have defined $\hat{\Sigma}_{z} = \hat{\sigma}_{1}^{z} + \sigma_{2}^{z}$.
Equation \eqref{Rotation Operator} shows that the constant of motion $\hat{\sigma}_{1}^{z} \hat{\sigma}_{2}^{z}$ is indeed a parity operator with respect to the collective spin-Pauli variable $\hat{\Sigma}_{z}$, since in correspondence to its integer eigenvalues $M = 0,\pm2$, $\hat{\sigma}_{1}^{z} \hat{\sigma}_{2}^{z}$ has eigenvalues $+1$ and $-1$ respectively.

It is important to underline that the existence of this constant of motion implies the existence of two sub-dynamics related to the two eigenvalues of $\hat{\sigma}_{1}^{z} \hat{\sigma}_{2}^{z}$. We can extract these two sub-dynamics by considering that the operator $\hat{\sigma}_{1}^{z} \hat{\sigma}_{2}^{z}$ has the same spectrum of $\hat{\sigma}_{2}^{z}$, i.e., the same eigenvalues ($\pm 1$) with the same twofold degeneracy. Therefore there exists a unitary time-independent operator $\mathbb{U}$ transforming $\hat{\sigma}_{1}^{z} \hat{\sigma}_{2}^{z}$ in $\hat{\sigma}_{2}^{z}$. It can be easily seen that the unitary and hermitian operator
\begin{equation}
\mathbb{U}=\dfrac{1}{2}\left[\openone+\hat{\sigma}_{1}^{z}+\hat{\sigma}_{2}^{x}-\hat{\sigma}_{1}^{z}\hat{\sigma}_{2}^{x}\right]=
\begin{pmatrix}
1 & 0 & 0 & 0 \\
0 & 1 & 0 & 0 \\
0 & 0 & 0 & 1 \\
0 & 0 & 1 & 0 \\
\end{pmatrix}
\end{equation}
in the standard ordered basis
\begin{equation}
\mathcal{B} = \{ \ket{++}, \ket{+-}, \ket{-+}, \ket{--} \}
\end{equation}
accomplishes the desired transformation:
\begin{equation}
\mathbb{U^{\dagger}}\hat{\sigma}_{1}^{z}\hat{\sigma}_{2}^{z}\mathbb{U}=\mathbb{U}\hat{\sigma}_{1}^{z}\hat{\sigma}_{2}^{z}\mathbb{U}=\hat{\sigma}_{2}^{z}.
\end{equation}
Transforming ${H}$ into ${\tilde{H}} = \mathbb{U}^{\dagger} {H} \mathbb{U}$, we get
\begin{equation} \label{H tilde}
\begin{aligned}
{\tilde{H}} = &
\hbar \omega_{1} \hat{\sigma}_{1}^{z} + \hbar \omega_{2} \hat{\sigma}_{1}^{z} \hat{\sigma}_{2}^{z} + \gamma_{zz} \hat{\sigma}_{2}^{z}\\
& + \gamma_{xx} \hat{\sigma}_{1}^{x} - \gamma_{yy} \hat{\sigma}_{1}^{x} \hat{\sigma}_{2}^{z} +
\gamma_{xy} \hat{\sigma}_{1}^{y} \hat{\sigma}_{2}^{z} + \gamma_{yx} \hat{\sigma}_{1}^{y}.
\end{aligned}
\end{equation}

It is easy to check that $\hat{\sigma}_{2}^{z}$ is a constant of motion of $\tilde{H}$ and that, consequently, $\tilde{H}$ may be represented as
\begin{equation}
\begin{aligned}
{\tilde{H}}  & = \sum_{\sigma_{2}^{z}} {\tilde{H}}_{\sigma_{2}^{z}} \ketbra{\sigma_{2}^{z}}{\sigma_{2}^{z}} = \\
& = {\tilde{H}}_{+} \otimes \ketbra{+}{+} \hspace{0.1cm} + \hspace{0.1cm} {\tilde{H}}_{-} \otimes \ketbra{-}{-}
\end{aligned}
\end{equation}
where 
\begin{equation}
\begin{aligned}
{\tilde{H}}_{\sigma_{2}^{z}} = & \gamma_{zz}\sigma_{2}^{z} +
\hbar\left(\omega_{1}+\omega_{2}\sigma_{2}^{z}\right)\hat{\sigma}_{1}^{z} + \\
& + \left(\gamma_{xx}-\gamma_{yy}\sigma_{2}^{z}\right) \hat{\sigma}_{1}^{x} + \left(\gamma_{xy}\sigma_{2}^{z}+\gamma_{yx}\right) \hat{\sigma}_{1}^{y}.
\end{aligned}
\end{equation}
This implies the existence of two ($\sigma_{2}^{z} = \pm 1$) sub-dynamics relative to a fictitious spin 1/2 immersed in different magnetic fields, each one possessing three components with the $z$ one only depending on time.

\subsection{Evolution operator in the presence of inhomogeneous time-dependent magnetic field}
If $\omega_{1}$ and $\omega_{2}$ were time independent, we would be able to solve exactly the time-evolution problem related to the Hamiltonian ${H}$. Indeed it is straightforward to find the eigenstates of ${\tilde{H}}$ as 
\begin{equation}
\ket{\tilde{\psi}} = \ket{\phi_{1i}}_{\sigma_{2}^{z}} \otimes \ket{\sigma_{2}^{z}}
\end{equation}
($i = 1,2$) where $\ket{\phi_{1i}}_{\pm 1}$ are the two eigenvectors of $\tilde{H}_{\pm}$, that is the two eigenvectors related to the sub-dynamics with $\sigma_{2}^{z} = \pm 1$. Through the relation
\begin{equation}
\ket{\psi} = \mathbb{U} \ket{\tilde{\psi}},
\end{equation}
we can in turn find the eigenvectors of ${H}$ and the time evolution of an arbitrary state of the two spins.

In the time-dependent case (when $\omega_{1}$ and $\omega_{2}$ depend on time), thanks to the fact that the unitary and hermitian operator $\mathbb{U}$ is time independent, we succeed, in view of the structure possessed by ${\tilde{H}}$ as given by Eq.~\eqref{H tilde}, in decoupling the time-dependent Schr\"odinger equation into two time-dependent Schr\"odinger equations of single spin 1/2. Therefore, we can construct the time-evolution operator of the whole dynamics of the two interacting spin 1/2's, starting from the construction of the two time evolution operators of the two sub-dynamics of single spin 1/2.
Indeed, starting from the initial time-dependent Schr\"odinger equation for the evolution operator $\mathcal{U}$ generated by ${H}$
\begin{equation}\label{Schroedinger eq}
i \hspace{0.05cm} \hbar \hspace{0.1cm} \dot{\mathcal{U}} = {H} \hspace{0.1cm} \mathcal{U},
\hspace{1cm}
\mathcal{U}(0) = \openone,
\end{equation}
we have, since $\dot{\mathbb{U}} = 0$, 
\begin{equation} \label{S. Equation U Tilde}
i \hspace{0.05cm} \hbar \hspace{0.1cm} \dot{\tilde{\mathcal{U}}} = {\tilde{H}} \hspace{0.1cm} \tilde{\mathcal{U}}
\hspace{1cm}
\tilde{\mathcal{U}}(0) = \openone, 
\end{equation}
where $\tilde{\mathcal{U}} \equiv \mathbb{U}^{\dagger} \mathcal{U} \mathbb{U}$.
If we search the time-evolution operator of ${\tilde{H}}$ in the form
\begin{equation} \label{Tilded Evolution Operator}
\begin{aligned}
\tilde{\mathcal{U}} = \tilde{\mathcal{U}}_{+} \otimes \ketbra{+}{+} \hspace{0.1cm} + \hspace{0.1cm} \tilde{\mathcal{U}}_{-} \otimes \ketbra{-}{-}
\end{aligned}
\end{equation}
the time-dependent Schr\"odinger equation for $\tilde{\mathcal{U}}$ in \eqref{S. Equation U Tilde} is converted into the following two Cauchy problems related to the sub-dynamics associated to ${\tilde{H}}_{+}$ and ${\tilde{H}}_{-}$
\begin{equation} \label{S. Eq. U1 and U2}
\left\{
\begin{aligned}
&i \hspace{0.05cm} \hbar \hspace{0.1cm} \dot{\tilde{\mathcal{U}}}_{+} = {\tilde{H}}_{+} \hspace{0.1cm} \tilde{\mathcal{U}}_{+},
\hspace{1cm}
\tilde{\mathcal{U}}_{+}(0) = \openone,
\\
&i \hspace{0.05cm} \hbar \hspace{0.1cm} \dot{\tilde{\mathcal{U}}}_{-} = {\tilde{H}}_{-} \hspace{0.1cm} \tilde{\mathcal{U}}_{-},
\hspace{1cm}
\tilde{\mathcal{U}}_{-}(0) = \openone.
\end{aligned}
\right.
\end{equation}
If we are able to solve these two single spin-1/2 time-dependent Schr\"odinger equations we then can construct easily the unique time-evolution operator of the entire ${\tilde{H}}$ as given in Eq.~\eqref{Tilded Evolution Operator} and then the time-evolution operator of the initial ${H}$ through the following relation
\begin{equation} \label{Relation U-Utilde}
\mathcal{U} = \mathbb{U} \hspace{0.05cm} \mathcal{\tilde{U}} \hspace{0.05cm} \mathbb{U}^{\dagger} = \mathbb{U} \hspace{0.05cm} \mathcal{\tilde{U}} \hspace{0.05cm} \mathbb{U}.
\end{equation}
 
The importance of this result consists in the possibility of applying the \emph{Messina-Nakazato} approach \cite{Mess-Nak} to each of the two sub-dynamics of single spin 1/2 to generate a class of time-dependent exactly solvable models whose Hamiltonian could be generally written as in \eqref{Hamiltonian}.
It is also important to stress that the procedure and the result are valid also if all the coupling constants $\gamma$ were time-dependent, too, besides $\omega_{1}$ and $\omega_{2}$.
To illustrate such a possibility we solve in detail the quantum dynamics of the two coupled spins taking advantage of some results reported in Ref.~\cite{Mess-Nak}.

The two-dimensional matrix of each sub-dynamics we got before can be written as follows
\begin{equation}
\tilde{H}_{\pm}=\tilde{H'}_{\pm} \pm \gamma_{zz} \openone 
=\begin{pmatrix}
\Omega_{\pm}(t) & \Gamma_{\pm} \\
\Gamma_{\pm}^{*} & -\Omega_{\pm}(t)
\end{pmatrix}
\pm \gamma_{zz} \openone,
\end{equation}
where we have put
\begin{equation} \label{Omega+- and Gamma+-}
\begin{aligned}
& \Omega_{\pm}(t) = \hbar ( \omega_{1}(t) \pm \omega_{2}(t) ), \\
& \Gamma_{\pm} = (\gamma_{xx} \mp \gamma_{yy}) - i (\pm \gamma_{xy} + \gamma_{yx}).
\end{aligned}
\end{equation}
Since, denoting by $\mathcal{E}_{\pm}$ the time-evolution operator generated by $\tilde{H}'_{\pm}$, the evolution operator generated by $\tilde{H}_{\pm}$ is simply given by
\begin{equation}
\tilde{\mathcal{U}}_{\pm} = e^{\pm i \gamma_{zz} t / \hbar} \mathcal{E}_{\pm},
\end{equation}
we will search directly the time-evolution operator $\mathcal{E}_{\pm}$ in accordance with the Ref.~\cite{Mess-Nak}. 
Following the example given in the section 3.3 of \cite{Mess-Nak} and considering the case when the transverse component of the magnetic field is constant (because in our context all the internal coupling coefficients are time independent), the time-evolution operator for each sub-dynamics of single spin 1/2 may be cast in the form
\begin{equation}
\mathcal{E}_{\pm} =
\begin{pmatrix}
|a_{\pm}|e^{i \phi_{a}^{\pm}} & |b_{\pm}|e^{i \phi_{b}^{\pm}} \\
-|b_{\pm}|e^{-i \phi_{b}^{\pm}} & |a_{\pm}|e^{-i \phi_{a}^{\pm}}
\end{pmatrix}
\end{equation}
where
\begin{equation}
|a_{\pm}|=\cos\biggl[{|\Gamma_{\pm}|\over\hbar}\int_0^t\cos \bigl[ \Theta_{\pm}(t') \bigr] dt' \biggr]
\end{equation}
and, since $|a_{\pm}|^2 + |b_{\pm}|^2 = 1$,
\begin{equation}
|b_{\pm}|=\sin\biggl[{|\Gamma_{\pm}|\over\hbar}\int_0^t\cos \bigl[ \Theta_{\pm}(t') \bigr] dt' \biggr]
\end{equation}
and
\begin{equation} \label{Phia and Phib}
\begin{aligned}
\phi_{a}^{\pm} = - \Bigl( \dfrac{\Theta_{\pm}}{2} + \mathcal{R}_{\pm} \Bigr), \quad 
\phi_{b}^{\pm} = - \dfrac{\Theta_{\pm}}{2} + \mathcal{R}_{\pm} - \dfrac{\pi}{2}
\end{aligned}
\end{equation}
with
\begin{equation}
\mathcal{R}_{\pm} = {|\Gamma_\pm| \over \hbar} \int_0^t { \sin\Theta_{\pm}\over\sin\bigl[{2|\Gamma_{\pm}|\over\hbar}\int_0^{t'}\cos\Theta_{\pm} dt''\bigr] } dt'.
\end{equation}
Here $\Theta_{\pm}(t)$ are arbitrary well-behaved mathematical functions fulfilling the condition $\Theta_\pm(0) = 0$.
Consistently, the longitudinal components of the effective magnetic fields of the two sub-dynamics vary over time such that
\begin{equation}
\Omega_{\pm}={\hbar\over2}\dot\Theta_{\pm}+|\Gamma_{\pm}|\sin\Theta_{\pm}\cot\biggl[{2|\Gamma_{\pm}|\over\hbar}\int_0^t\cos\Theta_{\pm} dt' \biggr]
\end{equation}
and we get easily the time dependence of $\omega_{1}$ and $\omega_{2}$ (and that of $B_1^z$ and $B_2^z$ through the relation \eqref{Relation omega-B}) resulting
\begin{equation} \label{omega1 and omega2 from Omega+ and Omega-}
\begin{aligned}
& \omega_{1} = \dfrac{\Omega_{+} + \Omega_{-}}{2 \hbar},
\quad
& \omega_{2} = \dfrac{\Omega_{+} - \Omega_{-}}{2 \hbar}.
\end{aligned}
\end{equation}
These equations practically single out the class of time-dependent Hamiltonians exactly treatable when the realistic assumption that the tensors $\mathbf{\Gamma}_{12}$, $\textbf{g}_{1}$ and $\textbf{g}_{2}$ are time independent is made.
Deriving the explicit form of the time-evolution operators related to the two independent sub-dynamics (described in terms of a single fictitious spin 1/2), we can write the whole unitary evolution operator of ${\tilde{H}}$ ($\tilde{\mathcal{U}}$) as prescribed in \eqref{Tilded Evolution Operator} and through relation \eqref{Relation U-Utilde} we get the unitary time-evolution operator of our initial problem, which reads
\begin{equation}\label{Total time ev op}
\mathcal{U} =
\begin{pmatrix}
|a_{+}|e^{i \Phi_{a}^{+}} & 0 & 0 & |b_{+}|e^{i \Phi_{b}^{+}} \\
0 & |a_{-}|e^{i \Phi_{a}^{-}} & |b_{-}|e^{i \Phi_{b}^{-}} & 0 \\
0 & -|b_{-}|e^{-i \Phi_{b}^{'-}} & |a_{-}|e^{-i \Phi_{a}^{'-}} & 0 \\
-|b_{+}|e^{-i \Phi_{b}^{'+}} & 0 & 0 & |a_{+}|e^{-i \Phi_{a}^{'+}} \\
\end{pmatrix},
\end{equation}
where we have put
\begin{subequations}\label{Big Phi definition}
\begin{align}
& \Phi_{a/b}^{\pm} = \phi_{a/b}^{\pm} \pm \dfrac{\gamma_{zz}}{\hbar} t \\
& \Phi_{a/b}^{' \pm} = \phi_{a/b}^{\pm} \mp \dfrac{\gamma_{zz}}{\hbar} t.
\end{align}
\end{subequations}

It is important to point out that if $\omega_{1}(t) = \omega_{2}(t)$ we have $\Omega_{-} = 0$ and the sub-dynamic related to $\sigma_{2}^{z} = -1$ is generated by
\begin{equation} \label{H minus static}
\hat{\tilde{H}}_{-} =
\begin{pmatrix}
0 & \Gamma_{-} \\
\Gamma_{-}^{*} & 0
\end{pmatrix}
\end{equation}
and then becomes a time-independent problem.
In this instance the relative evolution operator reads
\begin{equation} \label{Time Ev Op -1 Static Case}
\tilde{\mathcal{U}}_{-} = e^{-i {\gamma_{zz} \over \hbar} t}
\begin{pmatrix}
\cos \Bigl( \dfrac{|\Gamma_{-}|}{\hbar} t \Bigr) & e^{i \Phi} \sin \Bigl( \dfrac{|\Gamma_{-}|}{\hbar} t \Bigr) \\
e^{-i \Phi} \sin \Bigl( \dfrac{|\Gamma_{-}|}{\hbar} t \Bigr) & \cos \Bigl( \dfrac{|\Gamma_{-}|}{\hbar} t \Bigr)
\end{pmatrix},
\end{equation}
where $\Phi = \arctan \Bigl( {\gamma_{xx}+\gamma_{yy} \over \gamma_{yx}-\gamma_{xy}} \Bigr)$, and the whole evolution operator of the initial dynamic becomes
\begin{widetext}
\begin{equation} \label{Total Time Ev Op Static Case}
\begin{aligned}
\mathcal{U} =
\begin{pmatrix}
|a_{+}|e^{i \Phi_{a}^{+}} & 0 & 0 & |b_{+}|e^{i \Phi_{b}^{+}} \\
0 & e^{-i {\gamma_{zz} \over \hbar} t} \cos \Bigl( \dfrac{|\Gamma_{-}|}{\hbar} t \Bigr) & e^{i(\Phi-{\gamma_{zz} \over \hbar} t)} \sin \Bigl( \dfrac{|\Gamma_{-}|}{\hbar} t \Bigr) & 0 \\
0 & e^{-i(\Phi+{\gamma_{zz} \over \hbar} t)} \sin \Bigl( \dfrac{|\Gamma_{-}|}{\hbar} t \Bigr) & e^{-i {\gamma_{zz} \over \hbar} t} \cos \Bigl( \dfrac{|\Gamma_{-}|}{\hbar} t \Bigr) & 0 \\
-|b_{+}|e^{-i \Phi_{b}^{'+}} & 0 & 0 & |a_{+}|e^{-i \Phi_{a}^{'+}}
\end{pmatrix}.
\end{aligned}
\end{equation}
\end{widetext}
Of course, the evolution operator has the same form as that given by Eq.~\eqref{Total time ev op}, where the two-by-two internal block is now completely determined regardless of the way ${H}$ depends on time.
This means that when $\omega_{1}(t) = \omega_{2}(t) = \omega(t)$ in the Hamiltonian model given in Eq.~\eqref{Hamiltonian}, the time evolutions of $\ket{+-}$ and $\ket{-+}$ (and so, of every linear combination of these states) are independent of $\omega(t)$ and are characterized by Bohr frequencies related to the coupling constants appearing in ${H}$.

It is useful to underline that the condition $\omega_1(t)=\omega_2(t)$ is not implied simply by the condition $\mathbf{B}_1(t)=\mathbf{B}_2(t)$ because in general we may have different $\mathbf{g}$-tensors (or factors) for the two spins which ``rule'' the coupling with the magnetic field and are responsible for the different effective local magnetic fields in the two sites, even when $\mathbf{B}_1(t)=\mathbf{B}_2(t)$. So, the more general condition implying $\omega_1(t)=\omega_2(t)$ is
\begin{equation}\label{Condition equal omegas}
\mathbf{B}_1(t) \cdot \mathbf{g}_1 = \mathbf{B}_2(t) \cdot \mathbf{g}_2.
\end{equation}

\section{Exactly solvable time dependent scenarios for the two spin 1/2 model}\label{Section III}
In this section we report and discuss some particular time-dependent physical scenarios leading to exact analytical solutions of the Schr\"odinger equation \eqref{Schroedinger eq} by taking advantage of the approach described in the previous section. To this end we notice that on the basis of Eq.~\eqref{Tilded Evolution Operator} the knowledge of $\tilde{\mathcal{U}}_+$ and $\tilde{\mathcal{U}}_-$ is enough for determining the evolution operator $\mathcal{U}$ generated by the Hamiltonian ${H}$ governing the quantum dynamics of the two coupled spins. This means that in practice our task is the resolution of two dynamical problems each formally referred to a spin 1/2. It is right this point that makes of relevance the method in Ref.~\cite{Mess-Nak}.

The following two subsections report two novel exact solutions of the quantum dynamics of a spin 1/2 based on the method developed in Ref.~\cite{Mess-Nak}. In practice to single out a treatable scenario amounts at engineering the time-dependent magnetic field acting upon the spin 1/2. Both scenarios are useful in our problem meaning that each of them allows the selection of appropriate time-dependent exactly solvable models for ${\tilde{H}}_{+}$ and ${\tilde{H}}_{-}$. The last subsection is dedicated to the explicit construction of the two spin Hamiltonian models emerging from the intermediate steps leading to ${\tilde{H}}_{+}$ and ${\tilde{H}}_{-}$.

\subsection{First exactly solvable time-dependent scenario for one spin 1/2}\label{First time dependent scenario}
We may put
\begin{equation} \label{Position 1}
{|\Gamma|\over\hbar}\int_0^t\cos\Theta dt'={1 \over 2}\arcsin \bigl[ \tanh(\gamma t) \bigr], \quad \gamma={2|\Gamma| \over \hbar}.
\end{equation}
With this choice $|a| (|b|)$ goes from 1(0), at $t=0$, to $1/\sqrt{2} (1/\sqrt{2})$, as $t \rightarrow \infty$. Indeed we have
\begin{equation} \label{a and b case 1}
\begin{aligned}
& |a(t)| = \sqrt{\dfrac{ \cosh(\gamma t) + 1 }{2 \cosh(\gamma t)}}, \quad
& |b(t)| = \sqrt{\dfrac{ \cosh(\gamma t) - 1 }{2 \cosh(\gamma t)}}.
\end{aligned}
\end{equation}
Moreover, we have
\begin{subequations} \label{SinTheta e CosTheta decreasing a Position 1}
\begin{align}
&\cos \Theta(t) = \dfrac{1}{\cosh(\gamma t)} \\
&\sin \Theta(t)  = \tanh(\gamma t)
\end{align}
\end{subequations}
and the integral $\mathcal{R}$ is trivially integrated to yield
\begin{equation}
\mathcal{R} = {\gamma \over 2} t.
\end{equation}
From \eqref{SinTheta e CosTheta decreasing a Position 1} we derive
\begin{subequations} \label{Theta and dot Theta decreasing a}
\begin{align}
&\dot{\Theta} = \dfrac{\gamma}{\cosh(\gamma t)} \\
&\Theta = 2 \arctan \Bigl[ \tanh \Bigl( {\gamma \over 2} t \Bigr) \Bigr]
\end{align}
\end{subequations}
so that we get
\begin{subequations} \label{phia and phib case 1}
\begin{align}
& \phi_{a} = - \arctan \Bigl[ \tanh \Bigl( {\gamma \over 2} t \Bigr) \Bigr] - {\gamma \over 2} t \\
& \phi_{b} = - \arctan \Bigl[ \tanh \Bigl( {\gamma \over 2} t \Bigr) \Bigr] + {\gamma \over 2} t - {\pi \over 2}
\end{align}
\end{subequations}
and the longitudinal component of the magnetic field varies as
\begin{equation} \label{Omega 1/2}
\Omega = 2 |\Gamma| \dfrac{1}{\cosh(\gamma t)}.
\end{equation}
The plot of ${\Omega \over |\Gamma|}$ is shown in Fig.~1 as a function of $\tau_1={2|\Gamma| \over \hbar}t$.
\begin{figure}[h!]
\centering
{\includegraphics[width=\columnwidth]{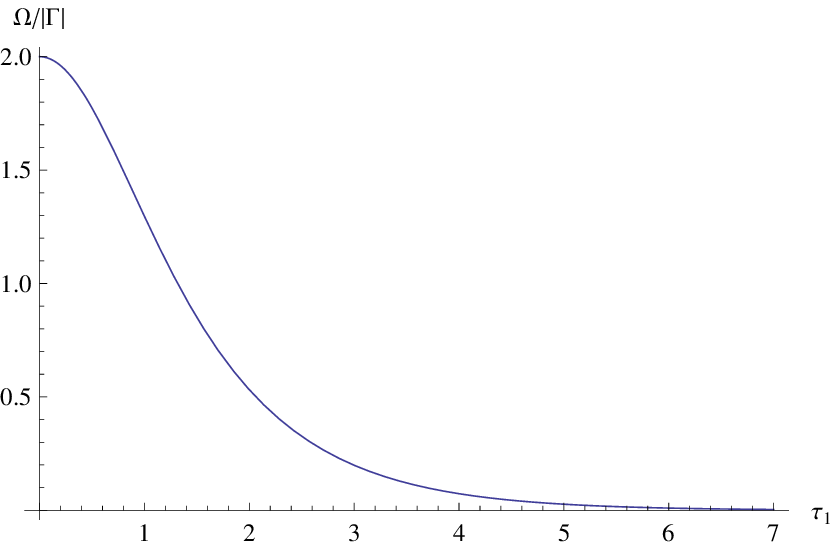} }
\caption{Plot of ${\Omega(\tau_1) \over |\Gamma|}$ according to Eq.~\eqref{Omega 1/2}.}
\end{figure}

\subsection{Second exactly solvable time-dependent scenario for one spin 1/2} \label{Second time dependent scenario}
We have a monotonically decreasing trend of the function $|a(t)|$ also by putting
\begin{equation} \label{Position 2}
{|\Gamma|\over\hbar}\int_0^t\cos\Theta dt'= \arcsin \bigl[ \tanh(\gamma t) \bigr], \quad \gamma={|\Gamma| \over \hbar},
\end{equation}
which, in view of Eqs.~(27) and (28), implies 
\begin{subequations} \label{a and b case 2}
\begin{align}
& |a(t)| = \dfrac{1}{\cosh(\gamma t)}, \\
& |b(t)| = \tanh(\gamma t).
\end{align}
\end{subequations}
In this case, thus, $|a|$ ($|b|$) varies from $1$($0$), at $t=0$, to $0$($1$) when $t \rightarrow \infty$, realizing a perfect inversion of the spin.
The expressions of $\cos\Theta(t)$ and $\sin\Theta(t)$ are the same as those in \eqref{SinTheta e CosTheta decreasing a Position 1} of the previous case and so also $\dot{\Theta}$ and $\Theta$ have the same expressions as those given in \eqref{Theta and dot Theta decreasing a} (though the definition of $\gamma$ is different in the two cases).
What is different is the value of integral $\mathcal{R}$  which, in this case, results in
\begin{equation}
\mathcal{R} = {1 \over 2} \sinh (\gamma t)
\end{equation}
and for the phases of $a$ and $b$ we have
\begin{subequations}\label{phia and phib case 2}
\begin{align}
& \phi_{a} = - \arctan \Bigl[ \tanh \Bigl( {\gamma \over 2} t \Bigr) \Bigr] - {1 \over 2} \sinh (\gamma t), \\
& \phi_{b} = - \arctan \Bigl[ \tanh \Bigl( {\gamma \over 2} t \Bigr) \Bigr] + {1 \over 2} \sinh (\gamma t) - {\pi \over 2}.
\end{align}
\end{subequations}

With this choice the longitudinal component of magnetic field  must be engineered as
\begin{equation} \label{Omega 1}
\Omega = {|\Gamma| \over 2} \biggl[ { 3 \over \cosh(\gamma t) } - \cosh(\gamma t) \biggr].
\end{equation}
Fig.~2 shows the behaviour of ${\Omega \over |\Gamma|}$ in this case against $\tau_2={|\Gamma| \over \hbar}t$.
\begin{figure}[h!]
\centering
{\includegraphics[width=\columnwidth]{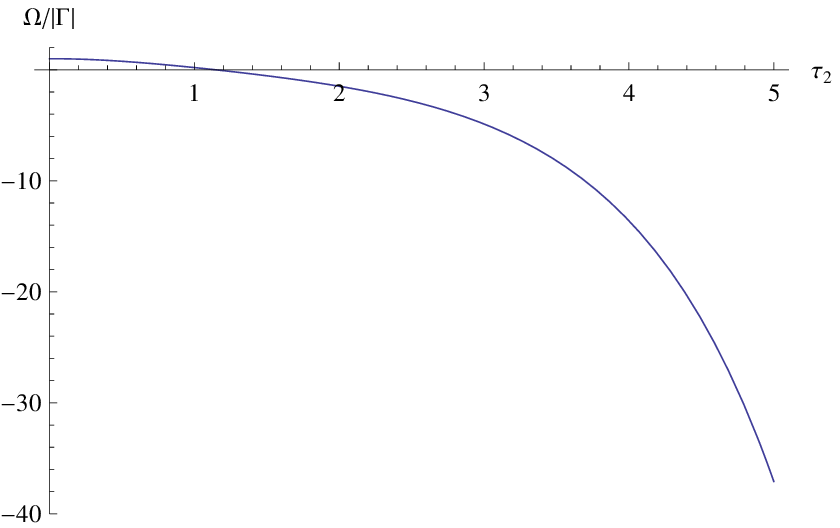} }
\caption{Plot of ${\Omega(\tau_2) \over |\Gamma|}$ according to Eq.~\eqref{Omega 1}.}
\end{figure}

It is important to point out that the value of the factor multiplying the function $\arcsin \bigl[\tanh(\gamma t) \bigr]$ is crucial for the possibility to solve exactly the integral $R$. Furthermore, it has a remarkable role in determining deeply the time evolution of important physical quantities as $|a|$, $|b|$ and $\Omega$. We saw, indeed, that the asymptotic ($t \rightarrow \infty$) values of $|a|$ and $|b|$ are very different in the two cases determining a completely different dynamical evolution in time. Finally, as we can see from Fig.~2, the multiplying factor significantly determines the time trend of the longitudinal component of magnetic field which must be engineered appropriately to have the exact dynamics we are studying.

\subsection{Time-dependent scenarios for the two spin model}
At closing this section, we emphasize the novelty of our results by explicitly giving all the time dependences (constructed  on the basis of Eq.~\eqref{omega1 and omega2 from Omega+ and Omega-}) of $\omega_1$ and $\omega_2$ (and so of the two magnetic fields $B_1^z$ ans $B_2^z$ in view of Eq.~\eqref{Relation omega-B}) in the two spin Hamiltonian model \eqref{Hamiltonian} leading to exactly solvable and solved models.
If we are interested in studying the time evolution of an initial state that belongs to one of the two dynamically invariant subspaces of ${H}$, wherein the dynamics is described by ${\tilde{H}}_{+}$ or ${\tilde{H}}_{-}$, we get classes of time dependent scenarios which can be treated and solved exactly. Precisely, if we consider, e.g., the sub-dynamics characterized by $\sigma_1^z\sigma_2^z=1$ and described by ${\tilde{H}}_{+}$, the two classes of time-dependent exactly solvable problems of two spins interacting according to our model in \eqref{Hamiltonian} are given by
\begin{subequations}\label{Relations B-Omega exactly scenario in first subspace}
\begin{align}
& \hbar \bigl(\omega_1(t)+\omega_2(t) \bigr)=\dfrac{2|\Gamma_+|}{\cosh({2|\Gamma_+|\over\hbar}t)}; \label{Relations B-Omega exactly scenario in first subspace a} \\
& \hbar \bigl(\omega_1(t)+\omega_2(t) \bigr)={|\Gamma_+| \over 2} \biggl[ { 3 \over \cosh({|\Gamma_+|\over\hbar}t)} - \cosh({|\Gamma_+|\over\hbar}t) \biggr]. \label{Relations B-Omega exactly scenario in first subspace b}
\end{align}
\end{subequations}
Equations \eqref{Relations B-Omega exactly scenario in first subspace} make clear the reason why we are talking of classes of time-dependent exactly solvable models.
Indeed we see that we have different possible choices of the two magnetic fields $B_1^z$ and $B_2^z$ such that their combination, in accordance to Eq.~\eqref{Relation omega-B}, satisfies one of the previous conditions, getting different time-dependent scenarios in which we are able to know exactly the dynamics. 
Obviously we have the analogous situation also for the other sub-dynamics characterized by $\sigma_1^z\sigma_2^z=-1$ and described by ${\tilde{H}}_{-}$. In this case the classes of exactly solvable models are due by the conditions
\begin{subequations}\label{Relations B-Omega exactly scenario in second subspace}
\begin{align}
& \hbar \bigl(\omega_1(t)-\omega_2(t) \bigr)=\dfrac{2|\Gamma_-|}{\cosh({2|\Gamma_-|\over\hbar}t)}; \label{Relations B-Omega exactly scenario in second subspace a} \\
& \hbar \bigl(\omega_1(t)-\omega_2(t) \bigr)={|\Gamma_-| \over 2} \biggl[ { 3 \over \cosh({|\Gamma_-|\over\hbar}t)} - \cosh({|\Gamma_-|\over\hbar}t) \biggr]. \label{Relations B-Omega exactly scenario in second subspace b}
\end{align}
\end{subequations}
We stress that Eqs.~\eqref{Relations B-Omega exactly scenario in second subspace} are not compatible with the situation corresponding to $\omega_1(t)=\omega_2(t)$ for which, on the other hand, the quantum dynamics in the subspace under scrutiny has been completely solved as explicitly given by Eq.~\eqref{Total Time Ev Op Static Case}. In other words, Eqs.~\eqref{Relations B-Omega exactly scenario in second subspace} display their usefulness, generating exactly solvable time-dependent Hamiltonian models of the two spins, only when $\omega_1(t) \neq \omega_2(t)$.
If we look, instead, at the entire dynamics of the two interacting spin 1/2's, considering a general initial condition belonging to the total four dimensional Hilbert space $\mathcal{H}$ we have the following four exactly solvable time dependent cases [$+(-)$ in $\pm$ corresponds to 1(2)]
\begin{subequations}\label{Exactly scenarios entire dynamics}
\begin{align}
&\hbar \omega_{1/2}(t) = \dfrac{|\Gamma_+|}{\cosh({2|\Gamma_+|\over\hbar}t)} \pm \dfrac{|\Gamma_-|}{\cosh({2|\Gamma_-|\over\hbar}t)}; \label{Exactly scenarios entire dynamics a} \\
&\hbar \omega_{1/2}(t) = \dfrac{|\Gamma_+|}{\cosh({2|\Gamma_+|\over\hbar}t)}\nonumber\\
&\hspace{1.5cm} \pm {|\Gamma_-| \over 4} \biggl[ { 3 \over \cosh({|\Gamma_-|\over\hbar}t) } - \cosh({|\Gamma_-|\over\hbar}t) \biggr];\label{Exactly scenarios entire dynamics b} \\
&\hbar \omega_{1/2}(t) = {|\Gamma_+| \over 4} \biggl[ { 3 \over \cosh({|\Gamma_+|\over\hbar}t) } - \cosh({|\Gamma_+\over\hbar}t) \biggr]\nonumber\\
&\hspace{1.5cm} \pm \dfrac{|\Gamma_-|}{\cosh({2|\Gamma_-|\over\hbar}t)};\label{Exactly scenarios entire dynamics c}\\
&\hbar \omega_{1/2}(t) = {|\Gamma_+| \over 4} \biggl[ { 3 \over \cosh({|\Gamma_+|\over\hbar}t) } - \cosh({|\Gamma_+|\over\hbar}t) \biggr]\\ \notag
& \hspace{1.5cm} \pm {|\Gamma_-| \over 4} \biggl[ { 3 \over \cosh({|\Gamma_-|\over\hbar}t) } - \cosh({|\Gamma_-|\over\hbar}t) \biggr].\label{Exactly scenarios entire dynamics d}
\end{align}
\end{subequations}
For example, if we consider the time-dependent scenario given by Eq.~\eqref{Exactly scenarios entire dynamics a} with a particular choice
\begin{equation}\label{Special choice of gammas}
\gamma_x = \gamma_y = 2 \gamma_{xy} = 2 \gamma_{yx} = c,
\end{equation}
we have, in view of Eqs.~\eqref{Omega+- and Gamma+-} and \eqref{Position 1},
\begin{equation}\label{Special Gamma_p/m and gamma_p/m}
\begin{aligned}
|\Gamma_+| = c, \quad
|\Gamma_-| = 2c \quad
\gamma_+ = {2c \over \hbar}, \quad
\gamma_- = {4c \over \hbar}
\end{aligned}
\end{equation}
and the time behaviour of ${\hbar\omega_1 \over c}$ and ${\hbar\omega_2 \over c}$ in terms of $\tau_c={ct \over \hbar}$ is seen in Fig.~3 ($\hbar = 1$).
\begin{figure}[h!]
\centering
{\includegraphics[width=\columnwidth]{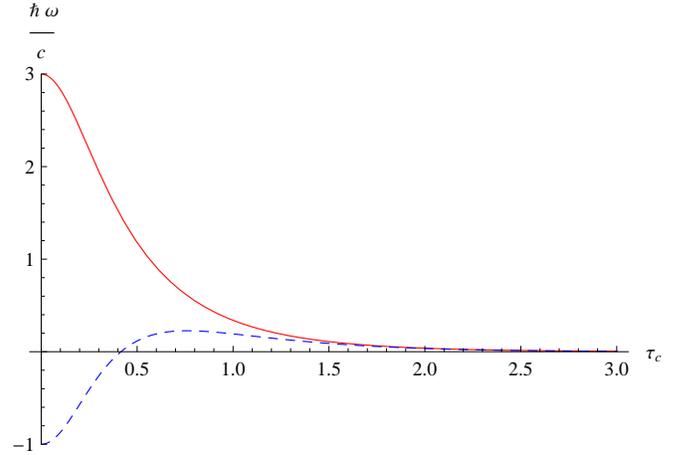} }
\caption{ Plots of ${\hbar\omega_1 \over c}$ (red solid line) and ${\hbar\omega_2 \over c}$ (blue dashed line), in terms of $\tau={ct \over \hbar}$, according to the time-dependent model satisfying Eqs.~\eqref{Exactly scenarios entire dynamics a} and \eqref{Special Gamma_p/m and gamma_p/m}. }
\end{figure}

The above four cases, so, provide $\omega_1(t)$ and $\omega_2(t)$ (and consequently the magnetic fields $B_1^z(t)$ and $B_2^z(t)$) such that our corresponding time-dependent Hamiltonian model given by Eq.~\eqref{Hamiltonian} turns out to be exactly solvable and the related global time-evolution operator $\mathcal{U}$, given by Eq.~\eqref{Total time ev op}, can be derived by plugging Eqs.~\eqref{a and b case 1} and \eqref{phia and phib case 1} or \eqref{a and b case 2} and \eqref{phia and phib case 2} in place of $\{ |a_+|,|b_+|,\phi_a^+,\phi_b^+ \}$ or of $\{ |a_-|,|b_-|,\phi_a^-,\phi_b^- \}$ at will, depending on what of the four time-dependent scenarios, given in Eqs.~\eqref{Exactly scenarios entire dynamics}, we choose.

We notice that when $g_1^{zz} \neq g_2^{zz}$, Eqs.~\eqref{Exactly scenarios entire dynamics}, by specializing Eq.~\eqref{Hamiltonian}, generate time-dependent Hamiltonian models of the two spins which cannot exactly be solved when a same external homogeneous magnetic field $B_1^z(t)=B_2^z(t)$ is applied on the two spins. In such a case, indeed, the time dependence of the magnetic field determined from Eq.~\eqref{Relations B-Omega exactly scenario in first subspace} is incompatible with that of the same magnetic field derivable from Eq.~\eqref{Relations B-Omega exactly scenario in second subspace}. However, it is of relevance to point out that, in this case, if we choose the time dependence of the unique magnetic field derived from either Eqs.~\eqref{Relations B-Omega exactly scenario in first subspace a} or \eqref{Relations B-Omega exactly scenario in first subspace b} $\bigl($\eqref{Relations B-Omega exactly scenario in second subspace a} or \eqref{Relations B-Omega exactly scenario in second subspace b}$\bigr)$, we get particular time-dependent Hamiltonian models for which we are able to solve exactly the sub-dynamics in the subspace singled out by the condition $\sigma_1^z\sigma_2^z=1$ $(\sigma_1^z\sigma_2^z=-1)$. It is finally useful to underline that when the physical system may be described assuming $g_1^{zz} = g_2^{zz}$, a homogeneous time-dependent magnetic field as derivable from either Eq.~\eqref{Relations B-Omega exactly scenario in first subspace a} or \eqref{Relations B-Omega exactly scenario in first subspace b}, leads to $\omega_1(t)=\omega_2(t)$ whose implications on the two spin quantum dynamics, have already been discussed after Eq.~\eqref{Big Phi definition}.

\section{Dynamical properties of the two spin 1/2 model} \label{Section IV}
In the previous sections we have built exactly solvable models for two coupled spin 1/2's and solved them as well. This result is important for two reasons. The first one is that it shows that the systematic route reported in Ref.~\cite{Mess-Nak} may be successfully applied to physical systems living in an $f$-dimensional Hilbert space with $f > 2$. The second one is related to the construction of new time-dependent exactly solvable Hamiltonian models in its own, since solutions of such problems, generally speaking, are very rare. Thus we are going to exploit the knowledge of the solutions we have found, in the subsections \ref{First time dependent scenario}, the first time-dependent scenario, and in \ref{Second time dependent scenario}, the second time-dependent scenario, to investigate physical properties exhibited by our two spin system under the corresponding engineered magnetic fields.

\subsection{Quantum evolution in the dynamically invariant subspace with parity $+$}
In the subspace where the constant of motion $\hat{S}_{1}^{z}\hat{S}_{2}^{z}$ assumes the value ${\hbar^2 \over 4}$ with certainty, in view of Eq.~\eqref{Total time ev op}, the initial states
\begin{equation}
\ket{\psi(0)} = \ket{\psi_{\alpha}^{+}(0)} \equiv \ket{++}
\end{equation}
and
\begin{equation}
\ket{\psi(0)} = \ket{\psi_{\beta}^{+}(0)} \equiv \ket{--}
\end{equation}
at time $t$, respectively, become

\begin{subequations} 
\begin{align}
&\ket{\psi_{\alpha}^{+}(t)} = |a_{+}|e^{i \Phi_{a}^{+}} \ket{++} - |b_{+}|e^{-i \Phi_{b}^{'+}} \ket{--}, \label{Time ev of ++} \\
&\ket{\psi_{\beta}^{+}(t)} = |b_{+}|e^{i \Phi_{b}^{+}} \ket{++} + |a_{+}|e^{-i \Phi_{a}^{'+}} \ket{--}. \label{Time ev of --}
\end{align}
\end{subequations}

Since both $\ket{++}$ and $\ket{--}$ are eigenvectors of $\hat{\mathbf S}^2$, the subspace they span pertains to the quantum number $S=1$ of such collective observable. It is worthwhile to observe that the magnetization $\average{\hat{S}^z(t)}_{\alpha/\beta}$ of the system in this subspace is not known with certainty.  Indeed, the mean value of $\hat{S}^{z} \equiv \hat{S}_{1}^{z} + \hat{S}_{2}^{z}$ on the states $\ket{\psi_{\alpha}^{+}(t)}$ and $\ket{\psi_{\beta}^{+}(t)}$, may be respectively cast as follows
\begin{subequations}\label{General time dependence of Sz}
\begin{align}
& \average{\hat{S}^z(t)}_\alpha \equiv \bra{\psi_{\alpha}^{+}(t)} \hat{S}^{z} \ket{\psi_{\alpha}^{+}(t)} = \hbar (|a_+|^2 - |b_+|^2), \\
& \average{\hat{S}^z(t)}_\beta \equiv \bra{\psi_{\beta}^{+}(t)} \hat{S}^{z} \ket{\psi_{\beta}^{+}(t)} = -\hbar (|a_+|^2 - |b_+|^2),
\end{align}
\end{subequations}
where $|a_+(t)|$ and $|b_+(t)|$ appear as entries of the matrix for $\mathcal{U}$.

\subsubsection{First time-dependent scenario}
As soon as $\omega_1(t)$ and $\omega_2(t)$ satisfy Eq.~\eqref{Relations B-Omega exactly scenario in first subspace a} we get the following magnetization time dependences
\begin{equation} \label{Sz2 position 1}
\begin{aligned}
\average{\hat{S}^z(t)}_{\alpha/\beta} = \pm {\hbar \over \cosh({2|\Gamma_+|\over\hbar}t)}.
\end{aligned}
\end{equation}
As a rule, the upper (lower) sign corresponds to $\alpha$ ($\beta$) here and in what follows.

It should be appreciated that the recipe provided by Eq.~\eqref{Relations B-Omega exactly scenario in first subspace a} enables in principle the construction of infinitely many time-dependent Hamiltonian models, all of them exactly predicting such an evolution of the magnetization of the coupled two spin system.
In Fig.~4 we plot such time dependences (for $\alpha$ and $\beta$) against the dimensionless time ${2|\Gamma_+|\over\hbar}t$.
\begin{figure}[h!]
\centering
{\includegraphics[width=\columnwidth]{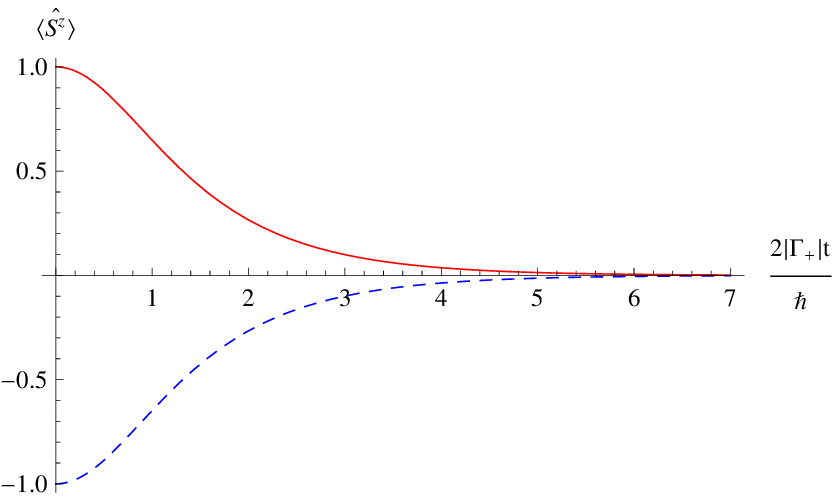} }
\caption{Time dependence of ${\average{\hat{S}^z(t)}_\alpha \over \hbar}$ (red solid line) and ${\average{\hat{S}^z(t)}_\beta \over \hbar}$ (blue dashed line) in the sub-dynamics with parity + for the class of time-dependent models characterized by Eq.~\eqref{Relations B-Omega exactly scenario in first subspace a}.}
\end{figure}

The common asymptotic value of $\average{\hat{S}^z(t)}_{\alpha}$ and $\average{\hat{S}^z(t)}_\beta$ can be understood by noticing that $\ket{\psi_{\alpha}^{+}(t)}$ and $\ket{\psi_{\beta}^{+}(t)}$, up to inessential phase factors, for large $t$ evolve into the following entangled Bell states of the two spins
\begin{subequations}\label{Asymptotic form of ++t and --t case 1}
\begin{align}
& \ket{\psi_{\alpha}^{+}(t)} \rightarrow e^{-i \bigl( {-\gamma_{zz} + |\Gamma_{+}| \over \hbar} t + {\pi \over 4} \bigr)} { \ket{++} + \ket{--} \over \sqrt{2} },\\
& \ket{\psi_{\beta}^{+}(t)} \rightarrow e^{i \bigl( {\gamma_{zz} + |\Gamma_{+}| \over \hbar} t + {3\pi \over 4} \bigr)} { \ket{++} - \ket{--} \over \sqrt{2} },
\end{align}
\end{subequations}
both having vanishing average magnetizations for $t\rightarrow \infty$. Equations \eqref{Asymptotic form of ++t and --t case 1} clearly evidence that engineering the magnetic fields on the two spins, in accordance with the constraint imposed by Eq.~\eqref{Relations B-Omega exactly scenario in first subspace a}, might be, in principle, at the heart of many possible experimental schemes successfully exploitable for generating such fully entangled states whatever the internal coupling coefficients appearing in the Hamiltonian model, given by Eq.~\eqref{Hamiltonian}, are.

\subsubsection{Second time-dependent scenario}
The time dependences of $\average{\hat{S}^z(t)}_{\alpha/\beta}$ in the sub-dynamics with parity + stemming from the class of time-dependent models characterized by Eq.~\eqref{Relations B-Omega exactly scenario in first subspace b} take, instead, the following forms
\begin{equation}
\begin{aligned}
\average{\hat{S}^z(t)}_{\alpha/\beta} = \pm \hbar \biggl( {2 \over \cosh^2({|\Gamma_+|\over\hbar}t)} - 1 \biggr),
\end{aligned}
\end{equation}
which is plotted in Fig.~5 as functions of ${|\Gamma_+|\over\hbar}t$.
\begin{figure}[h!]
\centering
{\includegraphics[width=\columnwidth]{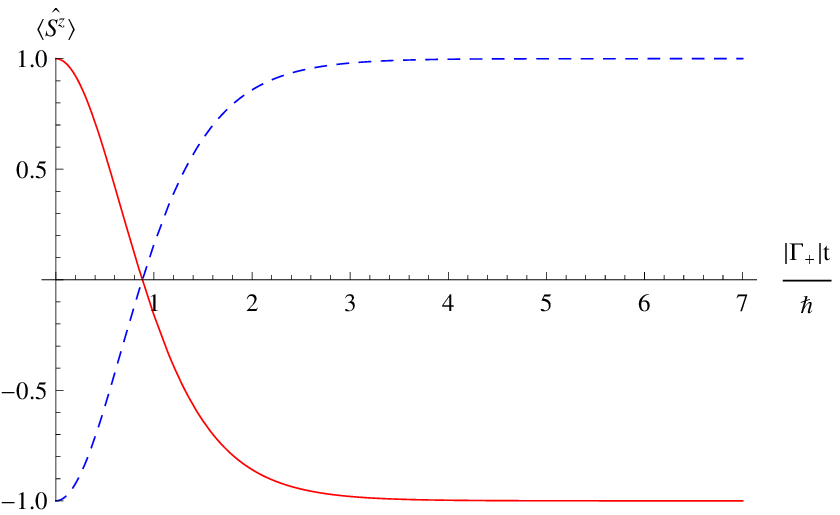} }
\caption{Time dependence of ${\average{\hat{S}^z(t)}_\alpha \over \hbar}$ (red solid line) and ${\average{\hat{S}^z(t)}_\beta \over \hbar}$ (blue dashed line) in the sub-dynamics with parity + for the class of time-dependent models characterized by Eq.~\eqref{Relations B-Omega exactly scenario in first subspace b}.}
\end{figure}

In this case a gradual inversion of the magnetization of the system occurs, due to the fact that as $t\rightarrow \infty$ we have a perfect inversion of the probability of finding the two spins in the state $\ket{++}$ ($\ket{--}$) when its initial state is $\ket{--}$ ($\ket{++}$). The asymptotic states for $t\rightarrow \infty$, in this scenario, are, indeed
\begin{equation}\label{Asymptotic form of ++t and --t case 2}
\begin{aligned}
& \ket{\psi_{\alpha}^{+}(t)} \rightarrow  e^{-i (\Phi_{b}^{'+} + \pi) } \ket{--}, \quad
& \ket{\psi_{\beta}^{+}(t)} \rightarrow  e^{i \Phi_{b}^{+}} \ket{++},
\end{aligned}
\end{equation}
implying immediately that
\begin{equation}
\Bigl|\average{\psi_{\alpha/\beta}^+(\infty)|\psi_{\alpha/\beta}^+(0)}\Bigr|^2 = 0.
\end{equation}

\subsection{Quantum evolution in the dynamically invariant subspace with parity $-$}
When $\hat{S}_{1}^{z}\hat{S}_{2}^{z}$ assumes the value $-{\hbar^2 \over 4}$ with certainty, the initial states
\begin{equation}
\ket{\psi(0)} = \ket{\psi_{\alpha}^{-}(0)} \equiv \ket{+-}
\end{equation}
and
\begin{equation}
\ket{\psi(0)} = \ket{\psi_{\beta}^{-}(0)} \equiv \ket{-+},
\end{equation}
evolve, respectively, as follows
\begin{subequations}
\begin{align}
&\ket{\psi_{\alpha}^{-}(t)} = |a_{-}|e^{i \Phi_{a}^{-}} \ket{+-} - |b_{-}|e^{-i \Phi_{b}^{'-}} \ket{-+}, \label{Time ev of +-}\\
&\ket{\psi_{\beta}^{-}(t)} = |b_{-}|e^{i \Phi_{b}^{-}} \ket{+-} + |a_{-}|e^{-i \Phi_{a}^{'-}} \ket{-+}. \label{Time ev of -+}
\end{align}
\end{subequations}
The states $\ket{+-}$ and $\ket{-+}$ generate the eigenspace of $\hat{S}^z$ of eigenvalue $M=0$, which, in turn, is not invariant with respect to the observable $\hat{\mathbf{S}}^2$, whose mean value runs from 0 to $2 \hbar^2$ in accordance with the following time evolutions:
\begin{equation} \label{S^2 Position 1}
\begin{aligned}
\average{\hat{\mathbf S}^{2}(t)}_{\alpha} & \equiv \bra{\psi_{\alpha}^{-}(t)} \hat{\mathbf S}^{2} \ket{\psi_{\alpha}^{-}(t)} = \\
& = \hbar^2 \Bigl[ 1 - 2 |a_-| |b_-| \cos(\phi_a + \phi_b) \Bigr], \\
\average{\hat{\mathbf S}^{2}(t)}_{\beta} & \equiv \bra{\psi_{\beta}^{-}(t)} \hat{\mathbf S}^{2} \ket{\psi_{\beta}^{-}(t)} = \\
& = \hbar^2 \Bigl[ 1 + 2 |a_-| |b_-| \cos(\phi_a + \phi_b) \Bigr].
\end{aligned}
\end{equation}

Equations \eqref{Relations B-Omega exactly scenario in second subspace} prescribe constraints on $\omega_1(t)$ and $\omega_2(t)$, with $\omega_1(t) \neq \omega_2(t)$, to guarantee the existence of exact dynamical behaviour of the corresponding models of the two spins, when the system is initially prepared in the subspace of parity $-$. 
When $\omega_1(t) = \omega_2(t)$, exploiting Eq.~\eqref{Total Time Ev Op Static Case}, we may easily get
\begin{equation}
\average{\hat{\mathbf S}^{2}(t)}_{\alpha/\beta} = \hbar^2 \biggl[ 1 \mp \sin \Bigl({2 |\Gamma_-| \over \hbar} t \Bigr) \cos(\Phi) \biggr],
\end{equation}
exhibiting oscillations at the Bohr frequency ${2 |\Gamma_-| \over \hbar}$ as expected on the basis of Eq.~\eqref{H minus static}.

In the next two subsections, $\omega_1(t) \neq \omega_2(t)$ is assumed.
\subsubsection{First time-dependent scenario}
The sub-dynamics with parity $-$ is exactly solvable for all the possible time-dependent models deducible from Eq.~\eqref{Relations B-Omega exactly scenario in second subspace a}. Accordingly, the time dependences of $\average{\hat{\mathbf S}^{2}}_{\alpha/\beta}$ turn out to be
\begin{equation}
\begin{aligned}
& \average{\hat{\mathbf S}^{2}(t)}_{\alpha/\beta} = \hbar^2 \Bigl[ 1 \pm \tanh^2({2|\Gamma_-|\over\hbar}t) \Bigr],
\end{aligned}
\end{equation}
which is graphically represented in Fig.~6 against the dimensionless time ${2|\Gamma_-|\over\hbar}t$.
\begin{figure}[h!]
\centering
{\includegraphics[width=\columnwidth]{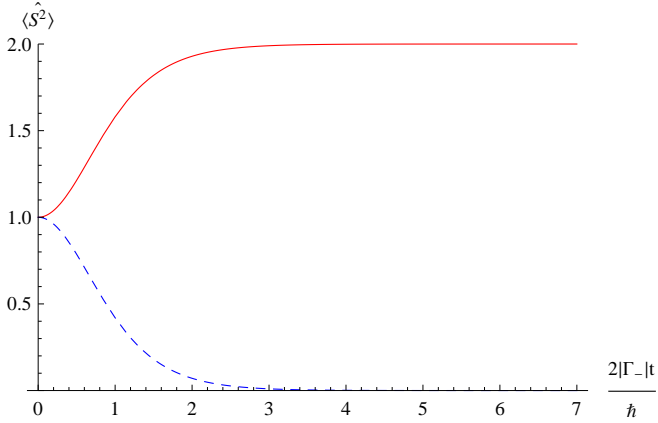} }
\caption{Time dependences of $\average{\hat{\mathbf S}^{2} (t)}_{\alpha}$ (red solid line) and $\average{\hat{\mathbf S}^{2} (t)}_{\beta}$ (blue dashed line) in the sub-dynamics with parity $-$ for the class of time-dependent models characterized by Eq.~\eqref{Relations B-Omega exactly scenario in second subspace a}.}
\end{figure}

The limiting values of $\average{\hat{\mathbf S}^2(t)}_{\alpha/\beta}$ for $t \rightarrow \infty$ suggest that the two spins asymptotically tend toward states identifiable as eigenstates of $\hat{\mathbf S}^2$ of eigenvalue $S=1$, $\ket{S=1,M=0}$, and $S=0$, $\ket{S=0,M=0}$, in correspondence to $\alpha$ and $\beta$, respectively. Thus, whatever $\omega_1(t) \neq \omega_2(t)$, fulfilling Eq.~\eqref{Relations B-Omega exactly scenario in second subspace a}, are, the envisioned time-dependent scenario under scrutiny leads to the generation of the following Bell states
\begin{subequations} \label{Asymptotic states second sub-dynamics first scenario}
\begin{align}
& \ket{\psi_{\alpha}^{-}(t)} \rightarrow e^{-i \bigl( {\gamma_{zz} + |\Gamma_{-}| \over \hbar} t + {\pi \over 4} \bigr)} { \ket{+-} + \ket{-+}  \over \sqrt{2} }\\
& \ket{\psi_{\beta}^{-}(t)} \rightarrow e^{i \bigl( {-\gamma_{zz} + |\Gamma_{-}| \over \hbar} t + {3\pi \over 4} \bigr)} { \ket{+-} - \ket{-+} \over \sqrt{2} }.
\end{align}
\end{subequations}
Even in this case, then, our results, as expressed by Eq.~\eqref{Asymptotic states second sub-dynamics first scenario}, might provide implementable experimental strategies for the generation of two maximally entangled states $\ket{S=1,M=0}$ and $\ket{S=0,M=0}$.

\subsubsection{Second time-dependent scenario}
All the time-dependent models satisfying Eq.~\eqref{Relations B-Omega exactly scenario in second subspace b} are characterized by time dependences of $\average{\hat{\mathbf S}^{2}}_{\alpha/\beta}$ in the sub-dynamics with parity $-$ of the following form
\begin{equation}
\begin{aligned}
& \average{\hat{\mathbf S}^{2}(t)}_{\alpha/\beta} = \hbar^2 \biggl[ 1 \pm { 2\tanh^2({|\Gamma_-|\over\hbar}t) \over \cosh({|\Gamma_-|\over\hbar}t) } \biggr],
\end{aligned}
\end{equation}
whose behaviour is illustrated in Fig.~7 against ${|\Gamma_-| \over \hbar}t$.
\begin{figure}[h!]
\centering
{\includegraphics[width=\columnwidth]{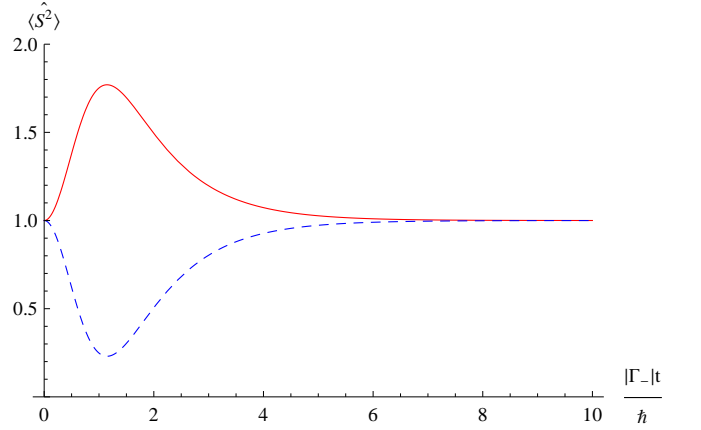} }
\caption{Time dependences of $\average{\hat{\mathbf S}^{2} (t)}_{\alpha}$ (red solid line) and $\average{\hat{\mathbf S}^{2} (t)}_{\beta}$ (blue dashed line) in the sub-dynamics with parity $-$ for the class of time-dependent models characterized by Eq.~\eqref{Relations B-Omega exactly scenario in second subspace b}.}
\end{figure}

Once again, the asymptotic trend of the curves reflects the $t \rightarrow \infty$ asymptotic states
\begin{equation}
\begin{aligned}
&\ket{\psi_{\alpha}^{-}(t)} \rightarrow  e^{-i (\Phi_{b}^{'-} + \pi)} \ket{-+}, \quad
&\ket{\psi_{\beta}^{-}(t)} \rightarrow  e^{i \Phi_{b}^{-}} \ket{+-},
\end{aligned}
\end{equation}
which means that under the strategy dictated by Eq.~\eqref{Relations B-Omega exactly scenario in second subspace b} the initial state $\ket{+-}$ is converted into the state $\ket{-+}$ while $\ket{-+}$ undergoes the analogous complete inversion.
\subsection{Quantum dynamics from an arbitrary initial condition}
In this last subsection we report the time evolution of a generic initial state in $\mathcal{H}$
\begin{equation}\label{Generic state in H}
\ket{\psi(0)} = c_{++} \ket{++} + c_{+-} \ket{+-} + c_{-+} \ket{-+} + c_{--} \ket{--}
\end{equation}
generated by one of the Hamiltonian models 
Eqs.~\eqref{Exactly scenarios entire dynamics}. Taking advantage of \eqref{Time ev of ++}, \eqref{Time ev of --}, \eqref{Time ev of +-} and \eqref{Time ev of -+}, we get formally
\begin{eqnarray}\label{Time evolved generic state in H}
\ket{\psi(t)} = && c_{++}(t) \ket{++} + c_{+-}(t) \ket{+-} + \\ \notag
&& + c_{-+}(t) \ket{-+} + c_{--}(t) \ket{--}
\end{eqnarray}
with
\begin{subequations} \label{Time dependence of the amplitudes for a generic state in H}
\begin{align}
& c_{++}(t) = e^{i{\gamma_{zz} \over \hbar}t} \Bigl( |a_+(t)|e^{i\phi_a^+(t)}c_{++}+|b_+(t)|e^{i\phi_b^+(t)}c_{--} \Bigr) \label{Time dependence of the amplitudes for a generic state in H a}\\
& c_{+-}(t) = e^{-i{\gamma_{zz} \over \hbar}t} \Bigl( |a_-(t)|e^{i\phi_a^-(t)}c_{+-}+|b_-(t)|e^{i\phi_b^-(t)}c_{-+} \Bigr) \\
& c_{-+}(t) = e^{-i{\gamma_{zz} \over \hbar}t} \Bigl( |a_-(t)|e^{-i\phi_a^{-}(t)}c_{-+}-|b_-(t)|e^{-i\phi_b^{-}(t)}c_{+-} \Bigr) \\
& c_{--}(t) = e^{i{\gamma_{zz} \over \hbar}t} \Bigl( |a_+(t)|e^{-i\phi_a^{+}(t)}c_{--}-|b_+(t)|e^{-i\phi_b^{+}(t)}c_{++} \Bigr), \label{Time dependence of the amplitudes for a generic state in H d}
\end{align}
\end{subequations}
where $|a_\pm(t)|$, $\phi_a^\pm(t)$, $|b_\pm(t)|$, $\phi_b^\pm(t)$ appear as entries in the matrix representation of the evolution operator $\mathcal{U}$ \eqref{Total time ev op} generated by the specific two spin Hamiltonian model under scrutiny (that is determined by one of \eqref{Exactly scenarios entire dynamics}).

The time evolutions of $\hat{\mathbf S}^{2}(t)$ as well as of $\hat{S}^z$ exhibit no interference terms stemming from the presence of states of different parity in \eqref{Time evolved generic state in H}. For example, in view of Eq.~\eqref{General time dependence of Sz}, we have
\begin{equation}
\average{\psi(t)|\hat{S}^z|\psi(t)} = \hbar (|c_{++}(t)|^2-|c_{--}(t)|^2)
\end{equation}
because the mean value of $\hat{S}^z$ in any state of negative parity identically vanishes.
It is thus interesting to evaluate the time evolution of the mean value of an observable which has nonvanishing matrix elements between states of different parities, for example $\hat{S}^x = \hat{S}_1^x+\hat{S}_2^x$. We limit ourselves to an exemplary case, namely the one obtained by choosing ${H}$ with $\omega_1(t)$ and $\omega_2(t)$ as prescribed in Eq.~\eqref{Exactly scenarios entire dynamics a} and the amplitudes of $\ket{\psi(0)}$ real and such to make the initial state a common eigenstate of $\hat{\mathbf{S}}^2$ and $\hat{S}^x$ with maximum eigenvalues, that is,
\begin{equation}
c_{++}=c_{+-}=c_{-+}=c_{--}={1 \over 2}.
\end{equation}
Equation \eqref{Time evolved generic state in H} immediately yields
\begin{widetext}
\begin{equation}
\begin{aligned}
\average{\hat{S}^x(t)} \equiv \average{\psi(t)|\hat{S}^x|\psi(t)} =
\hbar & \biggl[ \cos\Bigl( {2\gamma_{zz} \over \hbar} t \Bigr) \Bigl( |a_+||a_-|\cos(\phi_a^+)\cos(\phi_a^-)+|b_+||b_-|\sin(\phi_b^+)\sin(\phi_b^-) \Bigr)+\\
& + \sin\Bigl( {2\gamma_{zz} \over \hbar} t \Bigr) \Bigl( |a_+||b_-|\cos(\phi_a^+)\sin(\phi_b^-)-|a_-||b_+|\cos(\phi_a^-)\sin(\phi_b^+) \Bigr) \biggr]
\end{aligned}
\end{equation}
\end{widetext}
with
\begin{equation} \label{a b phia and phib scenario 53a}
\begin{aligned}
& |a_\pm(\tau_\pm)| = \sqrt{\dfrac{ \cosh(\tau_\pm) + 1 }{2 \cosh(\tau_\pm)}}, \\
& |b_\pm(\tau_\pm)| = \sqrt{\dfrac{ \cosh(\tau_\pm) - 1 }{2 \cosh(\tau_\pm)}}, \\
& \phi_a^\pm(\tau_\pm) = - \arctan \Bigl[ \tanh \Bigl( {\tau_\pm \over 2} \Bigr) \Bigr] - {\tau_\pm \over 2}, \\
& \phi_b^\pm(\tau_\pm) = - \arctan \Bigl[ \tanh \Bigl( {\tau_\pm \over 2} \Bigr) \Bigr] + {\tau_\pm \over 2} - {\pi \over 2},
\end{aligned}
\end{equation}
where $\tau_\pm = \gamma_\pm t$ and $\gamma_\pm = {2|\Gamma_\pm| \over \hbar}$, according to the time-dependent scenario \eqref{Exactly scenarios entire dynamics a} (that is, the ``first time-dependent scenario'' for each sub-dynamics). The plot of $\average{\hat{S}^x(t)}$, in this instance and considering the special case characterized by Eqs.~\eqref{Special choice of gammas} and \eqref{Special Gamma_p/m and gamma_p/m}, is given in Fig.~8 against $\tau_+$.
\begin{figure}[h!]
\centering
{\includegraphics[width=\columnwidth]{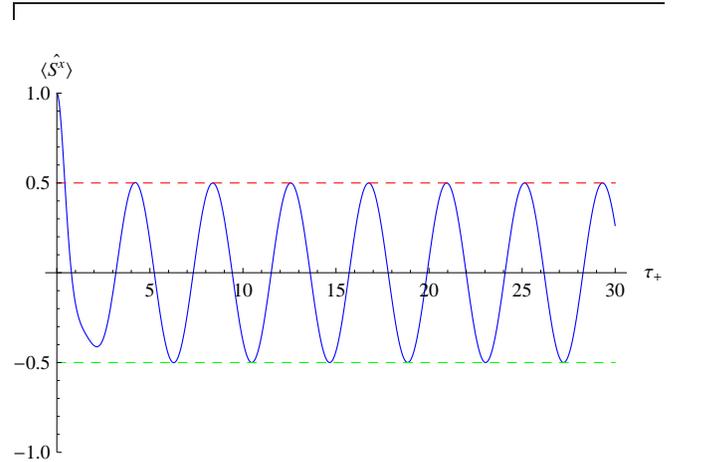} }
\caption{ Plot of $\average{\hat{S}^x(t)}$ (blue solid line) starting from $c_{++}=c_{+-}=c_{-+}=c_{--}={1 \over 2}$ according to the time-dependent scenario \eqref{Exactly scenarios entire dynamics a} and the special choice in Eqs. \eqref{Special choice of gammas} and \eqref{Special Gamma_p/m and gamma_p/m}; the red upper (green lower) dashed straight line represents $\average{\hat{S}^x(t)}={1 \over 2}$ ($\average{\hat{S}^x(t)}=-{1 \over 2}$). }
\end{figure}
It is possible to understand the peculiar behaviour for large $t$, characterized evidently by one frequency, by deriving analytically the asymptotic expression of $\average{\hat{S}^x(t)}$, which, indeed, acquires the following clear form
\begin{equation}
\average{\hat{S}^x(t)} = {\hbar \over 2} \cos \biggl[ \biggl( {\gamma_{zz} \over |\Gamma_+|} + {|\Gamma_-| \over 2 |\Gamma_+|} - {1 \over 2} \biggr) \tau_+ \biggr].
\end{equation}
\section{Concurrence in the two sub-dynamics} \label{Section V}
Spurred by the results of the previous section, which, in particular cases allow direct and first glance comparison between the initial level of entanglement with that get stored in the asymptotic states, in this section we are going to derive and analyse the exact time-evolution law of the entanglement established in the two spin system when it is initially prepared in the generic state given by Eq.~\eqref{Generic state in H}. For a pair of qubits, a good measure of entanglement is the concurrence $C$ introduced by Wooters \cite{Wootters} as well as the negativity, introduced by G. Vidal and R. F. Werner \cite{VidalWerner}, which in a generic state coincides with $C$ \cite{Grudka}. At a generic time instant $t$, it may be expressed as 
\begin{equation}\label{Concurence for a Generic State}
C(t) = 2 |c_{++}(t)c_{--}(t)-c_{+-}(t)c_{-+}(t)|,
\end{equation}
where the four time-dependent coefficients are the complex amplitudes of the normalized state $\ket{\psi(t)}$, into which $\ket{\psi(0)}$ evolves, and are given in Eqs.~\eqref{Time dependence of the amplitudes for a generic state in H}.
As expected, when the system starts in a state of definite parity, Eq.~\eqref{Concurence for a Generic State} yields $C(t)=2|c_{++}(t)c_{--}(t)|$ ($=2|c_{+-}(t)c_{-+}(t)|$) for parity $+$($-$).
When $c_{++}(0)=1$ or $c_{--}(0)=1$ and the first time-dependent scenario for this sub-dynamics is assumed, that is \eqref{Relations B-Omega exactly scenario in first subspace a}, the concurrence results
\begin{equation}\label{C starting from ++ first scenario}
C(t) = 2 |a(t)||b(t)| = \tanh({2|\Gamma_+|\over\hbar}t),
\end{equation}
whose plot is reported in Fig.~9 against $\tau_+={2|\Gamma_+|\over\hbar}t$.
\begin{figure}[h!]
\centering
{\includegraphics[width=\columnwidth]{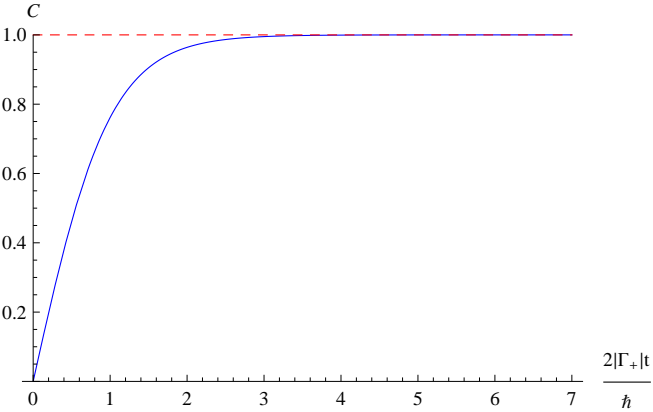} }
\caption{Plot of $C(t)$ starting from $\ket{++}$ when the time-dependent scenario \eqref{Relations B-Omega exactly scenario in first subspace a} is adopted.}
\end{figure}
The asymptotic behaviour of $C(t)$ in this case is easily understood in view of Eqs.~\eqref{Asymptotic form of ++t and --t case 1}.

Considering, instead, $c_{++}(0)=c_{--}(0)={1 \over \sqrt{2}}$, still together with \eqref{Relations B-Omega exactly scenario in first subspace a}, we obtain
\begin{eqnarray}\label{C starting from Bell state first scenario}
&&C(t) = \sqrt{ 1 - \tanh^2({2|\Gamma_+|\over\hbar}t) \sin^2({2|\Gamma_+|\over\hbar}t) },
\end{eqnarray}
which is plotted in Fig.~10 against $\tau_+={2|\Gamma_+|\over\hbar}t$.
\begin{figure}[H]
\centering
{\includegraphics[width=\columnwidth]{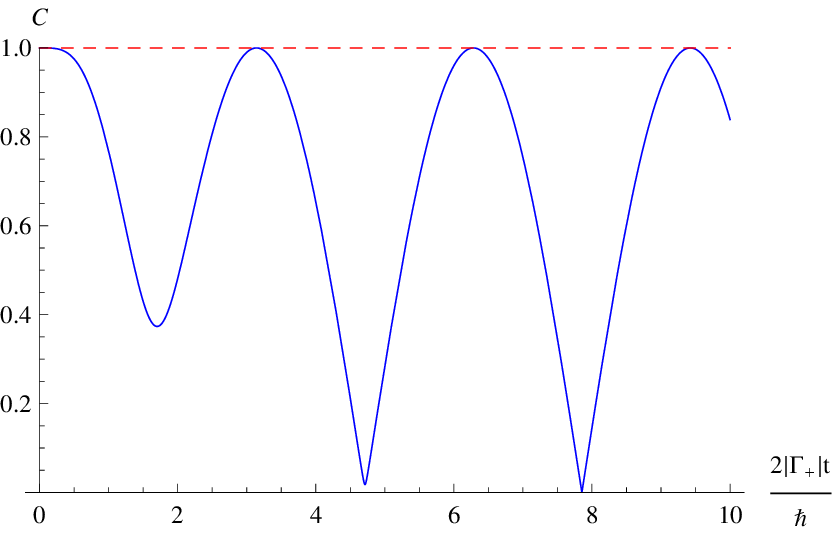} }
\caption{ Plot of $C(t)$ starting from ${\ket{++}+\ket{--} \over \sqrt{2}}$ under the time-dependent scenario \eqref{Relations B-Omega exactly scenario in first subspace a}. }
\end{figure}
It shows that, after a transient regime ($\tau_+ < 2\pi$), the concurrence oscillates between 0 and 1 as $|\cos(\tau_+)|$, which is immediately deduced from Eq.~\eqref{C starting from Bell state first scenario} for large $t$. The meaning of this behaviour is that the system periodically evolves alternating factorized states and the Bell states. To understand and better appreciate quantitatively this statement, we exploit Eqs. \eqref{Asymptotic form of ++t and --t case 1} to recover the asymptotic expression of $\ket{\psi(t)}$ ($\gamma_+={2|\Gamma_+|\over\hbar}$)
\begin{eqnarray}\label{Asymptotic state starting from Bell first scenario}
&&\ket{\psi(t)} \\ \notag
&& = e^{i {\gamma_{zz} \over \hbar} t} \Bigl[ e^{-i {\pi \over 2}}\sin\Bigl({\gamma_+ \over 2}t + {\pi \over 4}\Bigr) \ket{++} + \cos\Bigl({\gamma_+ \over 2} t + {\pi \over 4}\Bigr) \ket{--} \Bigr],
\end{eqnarray}
which clearly exhibits oscillations of period ${T}_+={4 \pi \over \gamma_+}={2\pi \hbar \over |\Gamma_+|}$ in accordance with the asymptotic expression of $C(t)$.
Equation \eqref{Asymptotic state starting from Bell first scenario} easily explains the oscillations exhibited by $C(t)$ since it predicts that the two spin system, up to a global phase factor, comes back to its initial condition and after a time ${{T}_+ \over 4}$ (${{T}_+ \over 2}$) it reaches the factorized (Bell-like) state $\ket{++}$ (${-i\ket{++}-\ket{--} \over \sqrt{2}}$) whereas in the last semi-period it reaches first the factorized state $\ket{--}$ and eventually its initial state.
The reason why the concurrence does not vanish in this case in the transient region stems from the fact that $C(t)=0$ necessarily implies $c_{++}(t)=0$ or $c_{--}(t)=0$. In view of the structure of the two amplitudes $c_{++}(t)$ and $c_{--}(t)$, see Eqs.~\eqref{Time dependence of the amplitudes for a generic state in H a} and \eqref{Time dependence of the amplitudes for a generic state in H d}, we deduce that $|a_+(t)|=|b_+(t)|$ is a necessary condition in order for $c_{++}(t)$ or $c_{--}(t)$ to vanish. Since such a condition is only asymptotically reached by the system, the concurrence cannot vanish during the transient regime.

We now study $C(t)$ when the system is initially prepared in one of the same three states considered above, adopting this time as Hamiltonian model the one stemming from Eq.~\eqref{Relations B-Omega exactly scenario in first subspace b} (called ``second time-dependent scenario'' in the previous section).
When $c_{++}(0)=1$ or $c_{--}(0)=1$ and the second time dependent scenario \eqref{Relations B-Omega exactly scenario in first subspace b}  is assumed, the concurrence becomes
\begin{equation}\label{C starting from ++ second scenario}
C(t)= 2{\tanh({|\Gamma_+|\over\hbar}t) \over \cosh({|\Gamma_+|\over\hbar}t)}
\end{equation}
and this is plotted in Fig.~11 as a fucntion of $\tau_+^{'}={|\Gamma_+|\over\hbar}t$.
\begin{figure}[h!]
\centering
{\includegraphics[width=\columnwidth]{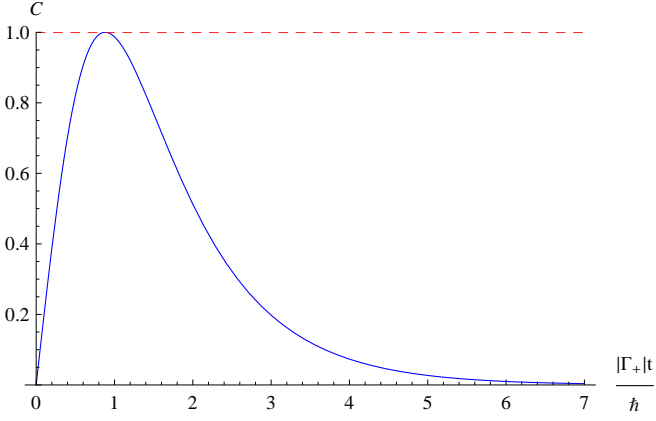} }
\caption{ Plot of $C(t)$ starting from $\ket{++}$ according to the time-dependent scenario \eqref{Relations B-Omega exactly scenario in first subspace b}. }
\end{figure}
We note that at time instant $(\tau_+)_0=\arcsinh(1)\approx 0.88$ the system of the two spins reaches maximally entangled state from which it asymptomatically evolves toward a factorized state. Even in this case it is useful to exploit Eqs.~\eqref{Time ev of ++} and \eqref{Time ev of --} together with Eqs.~\eqref{a and b case 2} and \eqref{phia and phib case 2} predicting that at the particular time instant $(\tau_+)_0$ the evolved states respectively become
\begin{subequations}
\begin{align}
& \ket{\psi_\alpha^+[(\tau_+)_0]}=e^{i\bigl({\gamma_{zz} (\tau_+)_0 \over \hbar} + \phi_a[(\tau_+)_0] \bigr)}{\ket{++}-e^{i({\pi \over 2}-1)} \ket{--} \over \sqrt{2}}, \\
& \ket{\psi_\beta^-[(\tau_+)_0]}=e^{i\bigl({\gamma_{zz} (\tau_+)_0 \over \hbar} + \phi_b[(\tau_+)_0] \bigr)}{\ket{++}+e^{-i({\pi \over 2}-1)} \ket{--} \over \sqrt{2}},
\end{align}
\end{subequations}
which is in accordance with the result on the concurrence.
Equations \eqref{Asymptotic form of ++t and --t case 2} provide the asymptotic form of the two evolutions under scrutiny, confirming the expectation of vanishing concurrence for large $t$.

If the two spin system is initially prepared in the Bell state $\ket{\psi(0)} = {\ket{++}+\ket{--} \over \sqrt{2}}$ the concurrence may be expressed as
\begin{equation}\label{C starting from Bell state second scenario}
C(t)= \sqrt{1-4{\tanh^2({|\Gamma_+|\over\hbar}t) \over \cosh^2({|\Gamma_+|\over\hbar}t)}\sin^2[\sinh({|\Gamma_+|\over\hbar}t)]},
\end{equation}
which is graphically represented in Fig.~12 as a function of $\tau_+^{'}={|\Gamma_+|\over\hbar}t$.
\begin{figure}[h!]
\centering
{\includegraphics[width=\columnwidth]{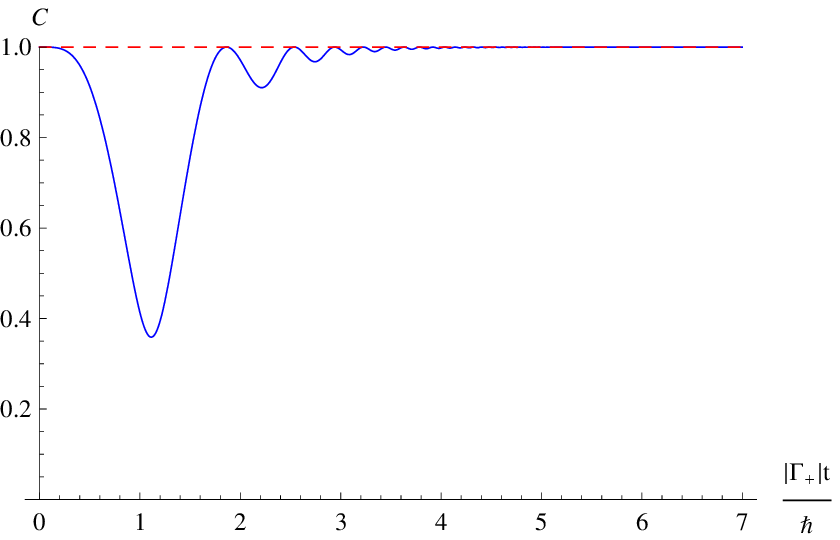} }
\caption{ Plot of $C(t)$ starting from ${\ket{++}+\ket{--} \over\sqrt2}$ according to the time-dependent scenario \eqref{Relations B-Omega exactly scenario in first subspace b}. }
\end{figure}
Since on the basis of Eqs.~\eqref{Asymptotic form of ++t and --t case 2} the state $\ket{++}$ ($\ket{--}$) asymptotically evolves into $\ket{--}$ ($\ket{++}$), up to an initial state-dependent global phase factor, the concurrence in the case under scrutiny must asymptotically comes back to its maximum value.
The peculiar oscillatory behaviour as time goes on, is due to the fact that the time evolution of $|c_{++}(t)|$ and $|c_{--}(t)|$ is dominated by progressive oscillations of decreasing amplitudes around $1/\sqrt{2}$ until they asymptotically stabilize at such values as we can transparently appreciate in Fig.~13.
\begin{figure}[h!]
\centering
{\includegraphics[width=\columnwidth]{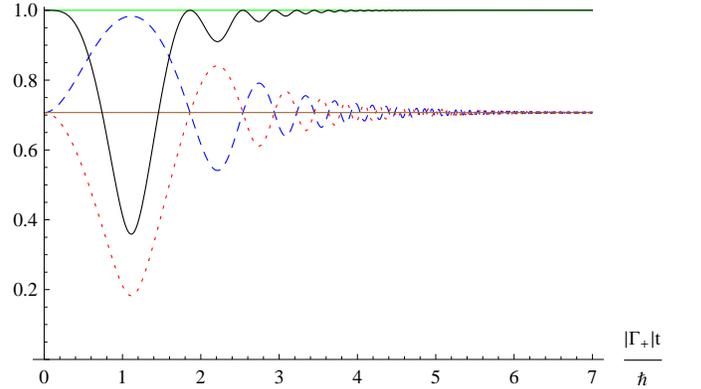} }
\caption{ Plots of $|c_{++}(t)|$ (blue dashed line) and $|c_{--}(t)|$ (red dotted line) and the resulting concurrence $C(t)$ (black line) starting from $c_{++}(0)={1 \over \sqrt{2}}$ and $c_{--}(0)={1 \over \sqrt{2}}$ according to the time-dependent scenario \eqref{Relations B-Omega exactly scenario in second subspace b}; the upper green (middle brown) line represents the $y=1$ ($y={1 \over \sqrt{2}}$) level.}
\end{figure}
The normalization of $\ket{\psi(t)}$ justifies the coincidence of the time instants where $|c_{++}(t)|$ and $|c_{--}(t)|$ assume the value $1/\sqrt{2}$. On the other hand the independence of $C(t)$ on the both phases of $c_{++}(t)$ and $c_{--}(t)$ explains why this time instants are exactly those at which $C(t)=1$. Moreover, Fig.~12 makes evident that all the infinitely-many minima of the concurrence occur at those time instants where $\Bigl||c_{++}(t)|-|c_{--}(t)|\Bigr|$ reaches local maxima in time (maximally unbalanced condition).

It is possible to find exactly the infinite sequence of states at which the concurrence assumes its maximum value. To this end we firstly calculate the time instants at which $C(t)=1$ and $|c_{++}(t)|=|c_{--}(t)|=1/\sqrt{2}$ simultaneously.
Thay are given by 
\begin{equation}
(\tau_+^{'})_n = \arcsinh(n\pi)
\end{equation}
with $n=0,1,2,\ldots$.
Plugging $(\tau_+^{'})_n$ into the state given in Eq.~\eqref{Time evolved generic state in H} after making explicit its time dependence with the help of Eqs.~\eqref{Time dependence of the amplitudes for a generic state in H a}, \eqref{Time dependence of the amplitudes for a generic state in H d} and \eqref{a b phia and phib scenario 53a} yields the following sequence of maximally entangled states progressively emerging in the time evolution of the initial Bell state:
\begin{equation}
\ket{\psi[(\tau_+^{'})_n]} = e^{i \bigl({\gamma_{zz} (\tau_+)_n \over \hbar} +\phi_n+\theta_n \bigr)} {\ket{++}+e^{-2i\phi_n}\ket{--} \over \sqrt{2}}
\end{equation} 
where
\begin{equation}\label{Phi_n}
\phi_n \equiv \phi_a^+[(\tau_+^{'})_n] = \sqrt{{\sqrt{1+(n\pi)^2}-1 \over \sqrt{1+(n\pi)^2}+1}}
\end{equation}
and
\begin{equation}
\theta_n = \arctan \Bigl( (-1)^{n+1} n\pi \Bigr),
\end{equation}
$n$ being an arbitrary nonnegative integer.

We emphasize that if we start from the initial condition $\ket{+-}$ adopting the time-dependent scenario \eqref{Relations B-Omega exactly scenario in second subspace a} (\eqref{Relations B-Omega exactly scenario in second subspace b}) we might again go through the arguments previously used to discuss the case $\ket{++}$ in conjunction with \eqref{Relations B-Omega exactly scenario in first subspace a} (\eqref{Relations B-Omega exactly scenario in first subspace b}), getting results that coincide with those expressed by Eqs.~\eqref{C starting from ++ first scenario} (\eqref{C starting from ++ second scenario}) provided that $\gamma_+$ is substituted by $\gamma_-$ and $\gamma_{zz}$ with $-\gamma_{zz}$. Analogously, had we started from the Bell state $(\ket{+-}+\ket{-+} )/\sqrt2$, Eq.~\eqref{C starting from Bell state first scenario} (\eqref{C starting from Bell state second scenario}) represents a valid result in this case too, provided the same substitution of $\gamma_+$ and $\gamma_{zz}$ are made and Eq.~\eqref{Relations B-Omega exactly scenario in second subspace a} (\eqref{Relations B-Omega exactly scenario in second subspace b}) is adopted.

It is worth noticing that in the parity-constrained dynamical evolution under scrutiny, the concurrence $C(t)$ at a generic time instant may be expressed as \cite{Palumbo}
\begin{equation}\label{Concurrence and Covariance}
C(t)=\sqrt{C_{xx}^2+C_{yy}^2}
\end{equation}
where
\begin{equation}
C_{xx}(t) =
\average{\psi(t)|\hat{\sigma}_1^x\hat{\sigma}_2^x|\psi(t)} - \average{\psi(t)|\hat{\sigma}_1^x|\psi(t)}\average{\psi(t)|\hat{\sigma}_2^x|\psi(t)}
\end{equation}
is the covariance of $\hat{\sigma}_1^x$ and $\hat{\sigma}_2^x$ and analogously $C_{yy}(t)$ is the covariance of $\hat{\sigma}_1^y$ and $\hat{\sigma}_2^y$. It is simple to show that, since \cite{Palumbo} $C_{xx}(t)=2Re[c_{++}(t)c^*_{--}(t)]$ and $C_{xx}(t)=2Im[c_{++}(t)c^*_{--}(t)]$, 
\begin{subequations}
\begin{align}
&C_{xx}[(\tau_+^{'})_n] = \cos(2 \phi_n) \\
&C_{yy}[(\tau_+^{'})_n] = \sin(2 \phi_n)
\end{align}
\end{subequations}
in accordance with the property $C[(\tau_+^{'})_n]=1$ for any $n=0,1,2,\ldots$.
Since, in view of Eq.~\eqref{Phi_n}, $2 \phi_n$ spans an infinite countable number set between 0 and 2 made up of irrationally related elements, $C_{xx}[(\tau_+^{'})_n]$ is a decreasing function of $n$, changing its sign as soon as $2(\tau_+^{'})_n > \pi$ and asymptotically tending to $\cos(2)$, whereas $C_{yy}[(\tau_+^{'})_n]$ is an increasing (decreasing) function of $n$ for $2 (\tau_+^{'})_n < \pi$ ($2 (\tau_+^{'})_n > \pi$), asymptotically tending to $\sin(2)$. It is remarkable that the quantitative link expressed by Eq.~\eqref{Concurrence and Covariance} enables a direct measurement of the level of the entanglement established in the system at any time instant.

\section{Conclusive Remarks} \label{Section VI}
The Hamiltonian model given by Eq.~\eqref{Hamiltonian} adopted in this paper contains seven parameters and then it is potentially useful to describe a huge variety of physical systems and/or physical situations in its parameter space. The key guidance leading us to extract this model from the general one given in Eq.~\eqref{General Hamiltonian} is the idea of assuring to our model the existence of a constant of motion with two eigenvalues only, holding at the same time the non-commutativity with $\hat{\mathbf{S}}^2$ and/or $\hat{S}^z$. Such a constant of motion, by construction, subdivides $\mathcal{H}$ into two dynamically invariant and orthogonal subspaces sharing the same dimension 2. The merit of such a decomposition is that it paves the way for extending our Hamiltonian model to a time-dependent scenario, namely that wherein the two spins are subjected to an appropriate inhomogeneous time-dependent magnetic field.

In this paper we report the exact time evolutions generated by such a time-dependent Hamiltonian.
This result is first of all important in its own since exact solvable problems involving two coupled bodies driven by time-dependent external fields are rare.
In connection with the last consideration, we point out that our exact treatment holds its validity even when the spin-spin coupling constants are time-dependent as for example happens when two neutral atoms located in the left and right sites of a double well are induced to merge in a single well by carefully adjusting (that is, time controlling) the trapping potential \cite{Anderlini}.
Our treatment possesses an additional merit of providing not a lucky trick confined to the problem under scrutiny only, but indeed an exportable route.
This claim on the one hand stems from the circumstance that the symmetry condition imposed to our Hamiltonian may be easily attributed to other Hamiltonian models representing dimers hosting two spins higher than /2 and even of different values.
On the other hand the consequent emergence of invariant sub-dynamics is traceable back to such a symmetry leading indeed to the possibility of taking advantage of the method reported in Ref.~\cite{Mess-Nak}.
Our treatment is illustrated finding the time behaviour of the two spins in correspondence to different choices of the inhomogeneous magnetic field.
In particular the time evolution of the mean value of some physically transparent observables as well as of the entanglement exhibited by the two qubit system during its time evolution is carefully reported and discussed.
Summing up, we wish to remark that providing exact solutions of a class of rather general Hamiltonian models describing two coupled qubits, although of relevance, is not the only result reported in this paper. We emphasize indeed that the strategic double exploitation of the decoupling treatment and of the systematic approach of Ref. \cite{Mess-Nak} demonstrates in a very transparent way the usefulness of such an approach beyond the original application to the quantum dynamics of a spin 1/2 subjected to a time-dependent magnetic field.
\section*{Acknowledgements}
A. M. and R. G. warmly thank G. Buscarino and M. Todaro for useful comments on possible experimental realizations of the Hamiltonian model and for suggesting Refs.~\cite{Napolitano} and \cite{Calvo}. Moreover, the same authors acknowledge stimulating conversation on the subject with B. Militello.


\begin{thebibliography}{99}
\bibitem{Bolton}
John A. Weil, James R. Bolton, Electron Paramagnetic Resonance - Elementary Theory and Practical Applications (Second Edition), John Wiley \& Sons (2007), Hoboken, New Jersey.
 
\bibitem{Thomas}
L. Thomas, F. Lionti, R. Ballou, D. Gatteshi, R. Sessoni and B. Barbara, Nature (London) \textbf{383}, 145 (1996).

\bibitem{Calbucci}
V. Calbucci, Ph.D. dissertation: Metodi matematici nelle scienze fisiche. Praxis: software per la simulazione del comportamento delle molecole magnetiche (2012).

\bibitem{Napolitano}
L. M. B. Napolitano, O. R. Nascimento, S. Cabaleiro, J. Castro and R. Calvo, Phys. Rev. B \textbf{77}, 214423 (2008).

\bibitem{Calvo}
R. Calvo, J. E. Abud, R. P. Sartoris and R. C. Santana, Phys. Rev. B \textbf{84}, 104433 (2011).

\bibitem{Bagrov}
M. C. Baldiotti, V. G. Bagrov and D. M. Gitman, Physics of Particles and Nuclei Letters, 2009, Vol. \textbf{6}, No. 7.

\bibitem{Anderlini}
M. Anderlini, P. J. Lee, B. L. Brown, J. Sebby-Strabley,2, W. D. Phillips and J. V. Porto, Nature \textbf{448}, 452-456 (2007).

\bibitem{Nguyen}
Van Hieu Nguyen, J. Phys.: Condens. Matter \textbf{21} (2009) 273201.

\bibitem{Imamoglu}
A. Imamoglu, D. D. Awschalom, G. Burkard, D. P. DiVincenzo, D. Loss, M. Sherwin, and A. Small, Phys. Rev. Lett. \textbf{83}, 4204 (1999).

\bibitem{Zheng}
Shi-Biao Zheng and Guang-Can Guo, Phys. Rev. Lett. \textbf{85}, 2392 (2000).

\bibitem{Wang}
Xiaoguang Wang, Phys. Rev. A \textbf{64}, 012313 (2001).

\bibitem{Arnesen}
M. C. Arnesen, S. Bose, and V. Vedral, Phys. Rev. Lett. \textbf{87}, 017901 (2001).

\bibitem{Yi}
X. X. Yi, H. T. Cui, and L. C. Wang, Phys. Rev. A \textbf{74}, 054102 (2006).

\bibitem{PorrasCirac}
D. Porras and J. I. Cirac, Phys. Rev. Lett. \textbf{92}, 207901 (2004).

\bibitem{Albayrak}
E. Albayrak, Eur. Phys. J. B \textbf{72}, 491496 (2009).

\bibitem{Guerrero}
R. J. Guerrero and M. F. Rojas, Quantum Inf Process \textbf{14}: 1973-1996 (2015).

\bibitem{Mess-Nak}
A. Messina and H. Nakazato, J. Phys. A: Math. Theor. \textbf{47}, 445302 (2014).

\bibitem{Wootters}
W. K. Wootters, Phys. Rev. Lett. \textbf{80}, 2245 (1998).

\bibitem{VidalWerner}
G. Vidal, R.F. Werner, Phys. Rev. A \textbf{65}, 032314 (2002).

\bibitem{Grudka}
A. Miranowicz and A. Grudka, Phys. Rev. A \textbf{70}, 032326 (2004).

\bibitem{Palumbo}
F. Palumbo, A. Napoli and A. Messina, Open Syst. Inf. Dyn. \textbf{13}, 309 (2006).

\end{thebibliography}
\end{document}